\documentclass[aps,pra,showkeys,twocolumn,superscriptaddress]{revtex4-2}
\usepackage{amsmath, amssymb}
\usepackage{mathrsfs}
\usepackage{array}
\usepackage{booktabs}
\usepackage{tabu}
\usepackage{dcolumn}
\usepackage{amsmath}
\usepackage{amsfonts}
\usepackage{float}
\usepackage{amssymb}
\usepackage{graphicx,color}
\usepackage[colorlinks={true}]
{hyperref}
\hypersetup{colorlinks=true,linkcolor=red,citecolor=blue,urlcolor=blue}
\usepackage{graphicx}
\usepackage{subfigure}
\usepackage{graphicx}
\usepackage{dcolumn}
\usepackage{bm}
\usepackage{pstricks}
\usepackage{braket}
\usepackage{orcidlink}

\def\be{\begin{equation}}
  \def\ee{\end{equation}}
\def\bea{\begin{eqnarray}}
\def\eea{\end{eqnarray}}
\def\f{\frac}
\def\n{\nonumber}
\def\l{\label}
\def\p{\phi}
\def\o{\over}
\def\R{\hat{\varrho}}
\def\pa{\partial}
\def\om{\Omega}
\def\na{\nabla}
\def\P{\Phi}
\begin{document}
\title{Ergotropy and capacity optimization in Heisenberg spin-chain quantum batteries
}

\author{Asad Ali\orcidlink{0000-0001-9243-417X}} \email{asal68826@hbku.edu.qa}
\affiliation{Qatar Centre for Quantum Computing, College of Science and Engineering, Hamad Bin Khalifa University, Doha, Qatar}
\author{Saif Al-Kuwari\orcidlink{0000-0002-4402-7710}} \email{smalkuwari@hbku.edu.qa}
\affiliation{Qatar Centre for Quantum Computing, College of Science and Engineering, Hamad Bin Khalifa University, Doha, Qatar}

\author{M. I. Hussain\orcidlink{0000-0002-6231-7746}}
\affiliation{Qatar Centre for Quantum Computing, College of Science and Engineering, Hamad Bin Khalifa University, Doha, Qatar}

\author{Tim Byrnes}
\affiliation{New York University Shanghai, NYU-ECNU Institute of Physics at NYU Shanghai, Shanghai Frontiers Science Center of Artificial Intelligence and Deep Learning, 567 West Yangsi Road, Shanghai, 200126, China}
 \affiliation{State Key Laboratory of Precision Spectroscopy, School of Physical and Material Sciences, East China Normal University, Shanghai 200062, China}
 \affiliation{Center for Quantum and Topological Systems (CQTS), NYUAD Research Institute, New York University Abu Dhabi, UAE}
 \affiliation{Department of Physics, New York University, New York, NY 10003, USA}

\author{M. T. Rahim\orcidlink{0000-0003-1529-928X}}
\affiliation{Qatar Centre for Quantum Computing, College of Science and Engineering, Hamad Bin Khalifa University, Doha, Qatar}

\author{James Q. Quach}
\affiliation{The University of Adelaide, SA 5005, Australia}

\author{Mehrdad Ghominejad\orcidlink{0000-0002-0136-7838}}
\affiliation{Faculty of Physics, Semnan University, P.O. Box 35195-363, Semnan, Iran}

\author{Saeed Haddadi\orcidlink{0000-0002-1596-0763}} \email{haddadi@semnan.ac.ir}
\affiliation{Faculty of Physics, Semnan University, P.O. Box 35195-363, Semnan, Iran}

\date{\today}
\def\be{\begin{equation}}
  \def\ee{\end{equation}}
\def\bea{\begin{eqnarray}}
\def\eea{\end{eqnarray}}
\def\f{\frac}
\def\n{\nonumber}
\def\l{\label}
\def\p{\phi}
\def\o{\over}
\def\R{\hat{\varrho}}
\def\pa{\partial}
\def\om{\Omega}
\def\na{\nabla}
\def\P{$\Phi$}

\begin{abstract}
This study examines the performance of finite spin \emph{quantum batteries} (QBs) using Heisenberg $XX$, $XY$, $XXZ$, and $XYZ$ spin models with Dzyaloshinsky-Moriya (DM) and Kaplan--Shekhtman--Entin-Wohlman--Aharony (KSEA) interactions. The QBs are modeled as interacting quantum spins in local inhomogeneous magnetic fields, inducing variable \emph{Zeeman splitting}. We derive analytical expressions for the maximal extractable work, \emph{ergotropy} and the \emph{capacity} of QBs, as recently examined by Yang \emph{et al.} [\href{https://doi.org/10.1103/PhysRevLett.131.030402}{Phys. Rev. Lett. 131, 030402 (2023)}]. These quantities are analytically linked through certain quantum correlations, as posited in the aforementioned study. Different Heisenberg spin chain models exhibit distinct behaviors under varying conditions, emphasizing the importance of model selection for optimizing QB performance. In antiferromagnetic (AFM) systems, maximum ergotropy occurs with a Zeeman splitting field applied to either spin, while ferromagnetic (FM) systems benefit from a uniform Zeeman field.  Temperature significantly impacts QB performance such that ergotropy in the AFM case is generally more robust against increasing temperature compared to the FM case. Incorporating DM and KSEA couplings can significantly enhance the capacity and ergotropy extraction of QBs. However, there exists a threshold beyond which additional increases in these interactions cause a sharp decline in capacity and ergotropy. This behavior is influenced by temperature and quantum coherence, which signal the occurrence of a sudden phase transition. The resource theory of quantum coherence proposed by Baumgratz \emph{et al.} [\href{https://doi.org/10.1103/PhysRevLett.113.140401}{Phys. Rev. Lett. 113, 140401 (2014)}] plays a crucial role in enhancing ergotropy and capacity. However, ergotropy is limited by both the system's capacity and the amount of coherence. These findings support the theoretical framework of spin-based QBs and may benefit future research on quantum energy storage devices.
\end{abstract}
\maketitle
\section{Introduction}\label{sec:1}
Ongoing battery research faces critical challenges, such as low energy density, slow charging speeds, limited shelf life, and environmental concerns. This introduces a pressing demand for innovative solutions to address these issues \cite{dell2001understanding,luo2015overview}.
The operation and architecture of chemical batteries are purely classical, completely disregarding the underlying useful quantum resource with the potential to leverage a new paradigm in the energy sector.
Recent theoretical proposals on \emph{quantum battery} (QB), based on the principles of quantum mechanics, present compelling heuristics \cite{ferraro2018high,campaioli2018quantum,binder2015quantacell}. { These proposals attempt to harness quantum traits such as superposition and quantum correlations \cite{shi2022entanglement,kamin2020entanglement,alicki2013entanglement,henao2018role} to tackle the challenges faced by conventional classical batteries, e.g., by offering superior energy densities and fast charging rates.}
This could potentially offer crucial insights into quantum thermodynamics, particularly in overcoming challenges related to environmental-induced superselection (einselection) in quantum mechanics \cite{albrecht2022einselection,zurek2003decoherence}. Recent findings also emphasize the importance of QBs for truly reversible quantum gates on quantum computers \cite{quach2023quantum}. Moreover, studying QBs can illuminate the quantum mechanical origins of thermodynamic quantities such as heat, internal energy, ergotropy, and entropy.

Although experimental realization of QBs in the literature is scarce,
{ a recent study \cite{hu2108optimal} reported experimental findings showing that a QB uses superconducting qutrits, which are optimized to maintain stable charging states.} Furthermore, the authors highlighted a self-discharge mechanism reminiscent of supercapacitors, suggesting efficient energy storage capabilities in superconducting circuits.
Similarly, recent experimental tests confirm a quantum advantage in charging QBs using NMR star-topology spin systems \cite{joshi2022experimental} and verified the charging behavior in organic QBs \cite{quach2020organic}.
Moreover, the authors in \cite{quach2022superabsorption} realized Dicke QBs, showcasing superextensive power storage capabilities in their experimental setup. More recently, QB charging utilizing single photons within a linear optics setup has been demonstrated \cite{huang2023demonstration}.
\begin{figure}[t]
    \centering
    \includegraphics[width=\columnwidth]{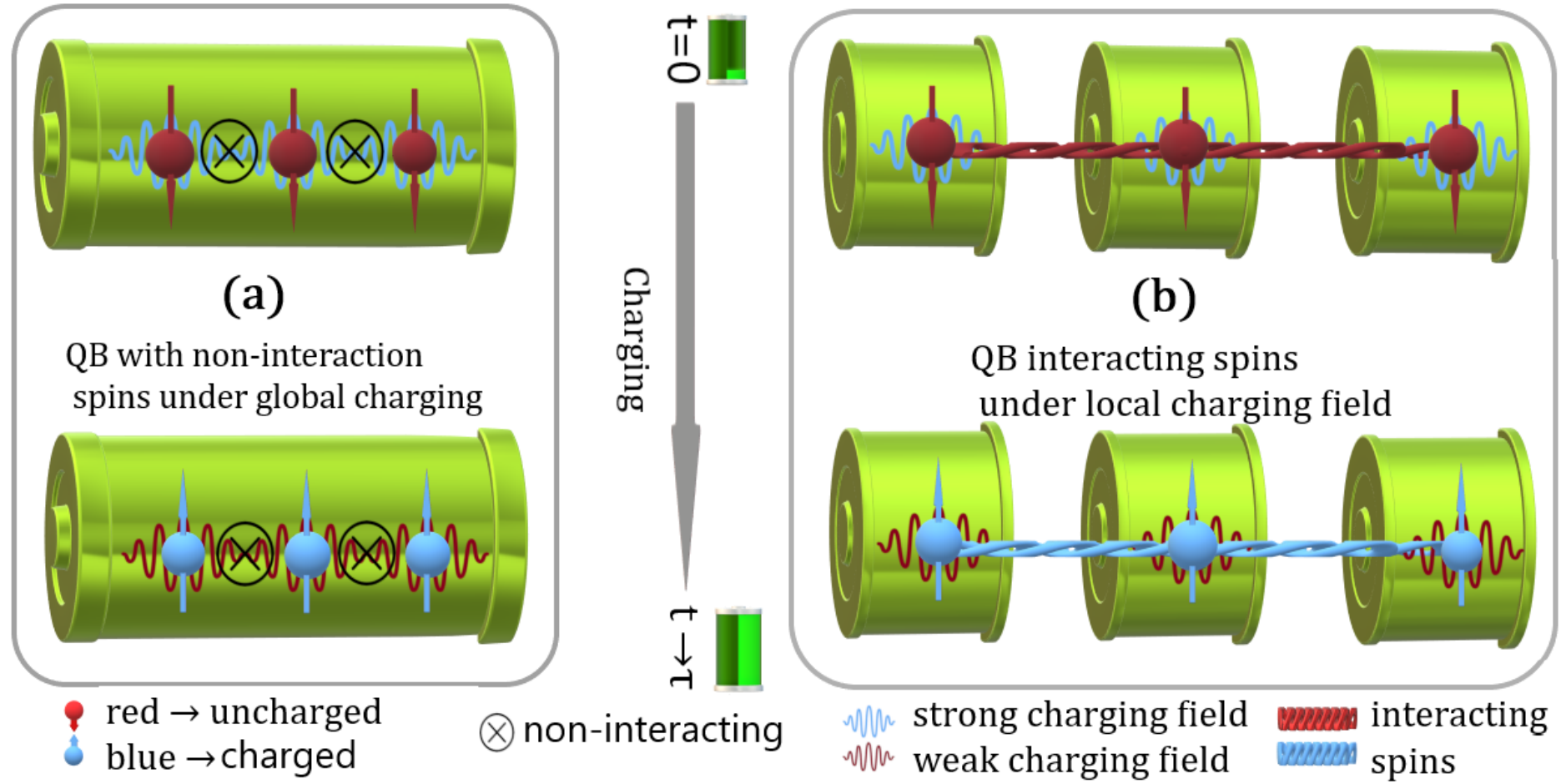}
    \caption{(a) Non-interacting QC-based QB with global charging field provided by the common cavity field without coupling between cells. (b) Interacting QC-based QB with local magnetic field provided by the field stored inside each cavity with couplings between cells.}
    \label{f1}
    \end{figure}

QBs are fundamentally realized through two distinct designs utilizing quantum cells (QCs) as their basic units \cite{le2018spin}, as depicted in Fig. \ref{f1}. The first design, the isolated QC-based QB, conceptualizes each QC as an independent, non-interacting entity, all subjected to a uniform global field for charging. These QCs only remain entangled during the charging process via a global entangled charger field, as shown in Fig. \ref{f1}(a). This approach primarily focuses on the intrinsic properties of each QC, such as quantum superposition and a non-degenerate energy spectrum. In contrast, the second design, the interacting QC-based QB, illustrated in Fig. \ref{f1}(b), considers each QC as part of a system where inter-QC couplings are pivotal in determining the QB’s overall performance in terms of energy storage and retrieval, acknowledging that no quantum system operates in isolation. This design evolves from isolated systems to a coupled many-body quantum system. QBs can be modeled using spin-chain models, which represent many-body quantum systems with couplings between constituent spins that function as QCs. Unlike non-interacting QC-based QBs, the interacting QC-based QB model necessitates at least two QCs or spins and emphasizes the importance of inter-spin couplings, as seen in one-dimensional spin chains with realistic two-body couplings \cite{grimaudo2018time,ali2024trade,grimaudo2016exactly,ali2024study,grimaudo2019coupling}.

Heisenberg spin chains (HSCs), which serve as foundational models in condensed matter physics and quantum information technologies, exhibit a variety of multipartite couplings and potential non-classical correlations such as quantum entanglement and quantum discord. These HSCs have been used extensively in quantum technologies including quantum transistors \cite{marchukov2016quantum}, cold atoms \cite{murmann2015antiferromagnetic}, interferometry \cite{de2008ramsey}, optical lattices \cite{simon2011quantum}, and quantum communications  \cite{bose2007quantum,yildirim1995anisotropic, lee2000entanglement, lee2013unconventional, gaudet2019quantum, patri2020distinguishing, signoles2021glassy, yeo2006teleportation, zhang2005thermal, ozaydin2015quantum, van2021quantum, cappellaro2007simulations, meier2003quantum, yu2023simulating, chepuri2023complex}. Since the introduction of the QB concept, researchers have been actively pursuing theoretical and experimental implementations across diverse quantum platforms, with the spin chain-based QB emerging as particularly promising \cite{dou2022cavity, yao2022optimal, ghosh2020enhancement, rossini2019many, mascarenhas2017nonequilibrium, huangfu2021high, arjmandi2023localization,  qi2021magnon, zakavati2021bounds, arjmandi2022enhancing, medina2023vqe,haseli2023,dwang2024,mojaveri2024extracting,PhysRevE.106.054107,PhysRevE.109.064103,lu2024topologicalquantumbatteries,PRXQuantum.5.030319,PhysRevLett.132.210402,Rodríguez_2024,Gyhm2024,Haseli2024RINP,haseli2024}. By leveraging spin chains as QBs, we can examine the interplay between exchange couplings and the charging process, potentially unlocking quantum advantages in terms of power and efficiency. Unlike earlier studies that focused on global entangling operations during charging, spin chain-based QBs facilitate an in-depth analysis of the role of local charging fields and the contribution of intra-system couplings to the charging process.

In this paper, { we focus on QBs based on the HSC with antisymmetric \emph{Dzyaloshinsky-Moriya} (DM) exchange coupling and symmetric \emph{Kaplan–
Shekhtman–Entin-Wohlman–Aharony} (KSEA) coupling} \cite{radhakrishnan2017quantum,radhakrishnan2017quantum2,ali2024study,ali2024trade} along with variable Zeeman splitting, which has not been addressed in the literature yet to the best of our knowledge. There are different sub-categories of 1D Heisenberg spin models, such as the Ising, $XX$, $XY$, $XXX$, $XXZ$, $XYZ$, and different research groups have addressed these specific cases of HSC models with intrinsic spin-spin couplings without DM and KSEA couplings \cite{dou2022cavity,zhao2021quantum,yao2022optimal,liu2024better,mojaveri2024extracting,mojaveri2024extracting,konar2024quantum,ghosh2020enhancement,grimaudo2016exactly}. Therefore, our major emphasis is on exploiting DM and KSEA couplings for extracting ergotropy production since in previous papers the spin-spin coupling in different subcategories of HSC such as Ising, $XX$, $XY$, $XXZ$, $XYZ$ have been reported. By exploring the spin-chain QB with added DM and KSEA couplings, we initiate a new avenue for experimentalists to design experimental schemes or working substances for spin chain-based efficient QBs, paving the way for the development of efficient and practical QBs tailored for quantum technologies.
The rest of this paper is organized as follows: Sec. \ref{sec:2} covers the modeling of QBs in a spin-$1/2$ chain, their initialization, and the charging process. It also introduces the performance indicators for QBs, including work done, ergotropy, power, and efficiency. Sec. \ref{sec:3} presents the results and discussion. Finally, Sec. \ref{sec:4} concludes this work.

\section{Modeling quantum batteries with Heisenberg spin Chain}\label{sec:2}
\subsection{Classes of Heisenberg spin models}
A spin-based QB is typically a system of quantum mechanical spins' collection with associated couplings to each other.
The general 1D Heisenberg spin-$1/2$ chain with nearest neighbor coupling can be modeled via a Hamiltonian, given by
\begin{align}
\hat{\mathcal{H}} =& \frac{1}{4}\sum_{j=1}^{N-1} J\bigl[(1+\gamma)\hat{\sigma}_j^x \otimes \hat{\sigma}_{j+1}^x
+ (1-\gamma)\hat{\sigma}_j^y \otimes \hat{\sigma}_{j+1}^y\bigr]\nonumber\\
&+ \frac{1}{4}\sum_{j=1}^{N-1} \Delta \hat{\sigma}_j^z \otimes \hat{\sigma}_{j+1}^z,
\label{eq1}
\end{align}
where $\gamma$ is called the anisotropy parameter, and $J$ and $\Delta$ are the nearest neighbor coupling parameters in the $xy$- and $z$-plane respectively. When both $J$ and $\Delta$ are positive, this model exhibits antiferromagnetic (AFM) behavior, whereas when both $J$ and $\Delta$ are negative, it demonstrates ferromagnetic (FM) characteristics. The standard Pauli spin operators denoted as $\hat{\sigma}_i^k$ (for $k = x, y, z$) define the spin vector on the { $i$th site} as $\overrightarrow{\mathbf{\mathcal{\sigma}}}_{i} = (\hat{\sigma}_i^x, \hat{\sigma}_i^y, \hat{\sigma}_i^z)$.
{ The different ranges of $\gamma$ and $\Delta$ determine how the model falls into different categories of the one-dimensional Heisenberg model, as shown in Table \ref{t1}.}

In this paper, we focus on four out of the six potential cases. Specifically, we analyze the $XX$, $XY$, $XXZ$, $XYZ$ models. This selection allows us to thoroughly investigate the key dynamics and characteristics pertinent to our research objectives.

For a more comprehensive understanding of the other cases not considered in this study, we direct the reader to the interesting work presented in \cite{ghosh2020enhancement,rossini2019many,le2018spin,rossini2019many,radhakrishnan2017quantum,radhakrishnan2017quantum,dou2022cavity,zakavati2021bounds,ullah2023low,dwang2024,mojaveri2024extracting}. We note that, to date, no research has incorporated DM and/or KSEA coupling(s) in the QBs based on various subclasses of HSCs.

\subsection{The QB Hamiltonian}
The Hamiltonian $\hat{\mathcal{H}}_{\mathcal{S}} $ for working substance of our system reads
\begin{equation}
\begin{split}
\hat{\mathcal{H}}_{\mathcal{S}} = \frac{1}{4}\sum_{j=1}^{N-1} \bigg\{J\bigl[(1+\gamma)\hat{\sigma}_j^x \otimes \hat{\sigma}_{j+1}^x  + (1-\gamma)\hat{\sigma}_j^y \otimes \hat{\sigma}_{j+1}^y\bigr]  \\+ \Delta\hat{\sigma}_j^z \otimes \hat{\sigma}_{j+1}^z +\overrightarrow{\mathbf{\mathcal{D}}}.(\overrightarrow{\mathbf{\mathcal{\hat{\sigma}}}}_{j}\times\overrightarrow{\mathbf{\mathcal{\hat{\sigma}}}}_{j+1})+ \overrightarrow{\mathbf{\mathcal{\hat{\sigma}}}}_{j}.\overleftrightarrow{\Gamma}.\overrightarrow{\mathbf{\mathcal{\hat{\sigma}}}}_{j+1}\bigg\},
\end{split}
\label{eq2}
\end{equation}
where the DM coupling is denoted by \(\overrightarrow{\mathbf{\mathcal{D}}}=(D_x,D_y,D_z)\) and the KSEA coupling $\overleftrightarrow{\Gamma}$ is a symmetric traceless tensor which can be written as
\begin{equation}
\overleftrightarrow{\Gamma} = \begin{pmatrix}
0 & G_{z} & G_{y} \\
G_{z} & 0 & G_{x} \\
G_{y} & G_{x} & 0 \\
\end{pmatrix}.
\label{eq3}
\end{equation}
The DM coupling, stemming from spin-orbit { interaction}, introduces an antisymmetric exchange coupling between neighboring spins, which is crucial for understanding phenomena like weak ferromagnetism \cite{shekhtman1992moriya,szilva2023quantitative,kuepferling2023measuring,zhang2022quantifying}. Simultaneously, the often-overlooked KSEA coupling restores symmetry disturbed by the DM term, providing a more accurate depiction of spin dynamics \cite{kaplan1983single}. Neglecting these couplings would obscure crucial aspects of quantum coherence and correlation, hindering our understanding of spin systems under external magnetic fields. Thus, their inclusion forms a fundamental basis for advancing QB technologies and deepening our understanding of correlated quantum materials-based QBs where DM and KSEA couplings are dominant \cite{Yurischev2023,Yurischev2020}.

{ In order to guarantee the non-degeneracy in the energy state of the working substance to act as QB, we apply Zeeman splitting to the spins using the following Hamiltonian
\begin{equation}
\hat{\mathcal{H}}_{\mathcal{Z}} = \sum_{j=1}^{N-1} \left[ \overrightarrow{\mathbf{\mathcal{B}}}_{j}  (\overrightarrow{\mathbf{\mathcal{\hat{\sigma}}}}_{j} \otimes \mathbb{\hat{I}}_2) + \overrightarrow{\mathbf{\mathcal{B}}}_{j+1}  (\mathbb{\hat{I}}_2  \otimes \overrightarrow{\mathbf{\mathcal{\hat{\sigma}}}}_{j+1}) \right],
\label{eq4}
\end{equation}
where $\overrightarrow{\mathbf{\mathcal{B}}}_{j}=(B_{j}^x,B_{j}^y,B_{j}^z)$ represents the magnetic field applied in an arbitrary direction to induce Zeeman splitting at any $j$th and $(j+1)$th site. We consider inhomogeneous magnetic fields along the chain, necessitating a site-dependent Hamiltonian formulation to capture this variation accurately. This approach is crucial for comprehensively understanding the system's dynamics and behaviors.}

\begin{table}[t]
\centering
\caption{Different cases of 1D Heisenberg spin models.}
\label{t1}
\begin{ruledtabular}
\begin{tabular}{ccc}
Anisotropy in $xy$-plane ($\gamma$) & $z$-axis coupling ($\Delta$) & Case \\
\colrule
$\pm1$ & $0$ & Ising  \\
$0$ & $0$ & $XX$  \\
$0$ & $\neq 0$ & $XXZ$  \\
$0$ & $J$ & $XXX$  \\
$0 < \gamma < 1$ & $0$ & $XY$  \\
$0 < \gamma < 1$ & $\neq 0$ & $XYZ$  \\
\end{tabular}
\end{ruledtabular}
\end{table}
{ Having the Hamiltonian of the working substance  \eqref{eq2} and the Zemaan splitting Hamiltonian  \eqref{eq4}, we can establish the Hamiltonian for a QB as}
\begin{equation}
\hat{\mathcal{H}}_{\mathcal{Q}}=\hat{\mathcal{H}}_{\mathcal{S}}+\hat{\mathcal{H}}_{\mathcal{Z}}.
\label{eq5}
\end{equation}

Given the intricate nature of the model, characterized by many parameters, maintaining complete generality becomes daunting. We add that not all subcategories of Heisenberg spin models are exactly solvable such that the analytical spectrum of energies and eigenstates is known. Thus, we opt for a tractable approach, focusing on a scenario involving two neighboring spins. { Within this framework, we set the magnitude of the magnetic field along $x$ and $y$ directions to be zero and investigate the influence of a Zeeman splitting due to an inhomogeneous sinusoidal magnetic field acting along the conventional $z$-direction, denoted as $\overrightarrow{\mathbf{\mathcal{B}}}_{1}=(0,0, B \cos{\theta})$ and $\overrightarrow{\mathbf{\mathcal{B}}}_{2}=(0,0, B\sin{\theta})$.}  Let us define $0\leq \theta \leq \frac{\pi}{2}$ such that when $\theta=0$ (or equivalently $\theta=\frac{\pi}{2}$ ), the Zeeman splitting is applied to only one spin. When $\theta=\frac{\pi}{4}$, the magnetic field is applied equally to both spins, and the same amount of Zeeman splitting is applied to both QCs. Furthermore, we consider the DM and KSEA couplings { to be aligned} with the $z$-direction, specified by $\overrightarrow{\mathbf{\mathcal{D}}}=(0,0, D_z)$, with the constraint $G_x=G_y=0$ and $G_z\neq0$. { Consequently, Eq. \eqref{eq5} through Eqs. \eqref{eq2} and \eqref{eq4} can be written to accommodate a pair of two spins, namely}

\begin{equation}
\begin{split}
\mathcal{\hat{H}}_{\mathcal{Q}} &= \frac{J}{4}[(1+\gamma)\hat{\sigma}_1^x \otimes \hat{\sigma}_{2}^x  + (1-\gamma)\hat{\sigma}_1^y \otimes \hat{\sigma}_{2}^y\bigr] \\
&+\Delta\hat{\sigma}_1^z \otimes \hat{\sigma}_{2}^z+ B \cos{\theta}(\mathbf{\mathcal{\hat{\sigma}}}_{1}^z\otimes\mathbb{\hat{I}}_2) + B \sin{\theta}(\mathbb{\hat{I}}_2\otimes\mathbf{\mathcal{\hat{\sigma}}}_{2}^{z}) \\
&+ D_z(\hat{\sigma}_1^x\otimes\hat{\sigma}_{2}^y -\hat{\sigma}_{1}^y\otimes\hat{\sigma}_{2}^x) + G_z (\hat{\sigma}_1^x\otimes\hat{\sigma}_{2}^y + \hat{\sigma}_{1}^y\otimes\hat{\sigma}_{2}^x).
\end{split}
\label{eq6}
\end{equation}
\subsection{The initial state of QB}
In the context of QBs, we may consider two initial states: the uncharged state and the charged state. For the uncharged QB's initial condition, it can either be in its ground state $|00\rangle\langle 00|$, representing absolute zero temperature, or in a thermal state with finite temperature \cite{ghosh2020enhancement,ullah2021characterization}. In the latter case, the system is initially in its ground state, and through a unitary operator, it undergoes a transition to an excited state, effectively fully charging the battery over time for subsequent power or ergotropy extraction. { Since in the real world no system can reach absolute zero, as the third law of thermodynamics and the Heisenberg uncertainty principle do not allow, we choose the second option and find Gibbs thermal state at equilibrium temperature $T$ from the Hamiltonian expressed in Eq. \eqref{eq6}. Hence, we get}
\begin{equation}
\hat{\varrho}_{th} = \frac{e^{-\hat{\mathcal{H}}_{\mathcal{Q}}/k_{B}T}}{Z}= \frac{1}{Z}\sum_{\mu=1}^{4} \exp\left(-\nu_\mu/k_{B}T\right) \ket{\psi_{\mu}}\bra{\psi_{\mu}}.
\label{eq10}
\end{equation}
In the above expression, $Z = \text{Tr}[\exp(-\hat{\mathcal{H}}_{\mathcal{Q}}/k_{B}T)]$ is the partition function and $\nu_\mu$ stands for the eigenenergy corresponding to the eigenstate $\ket{\psi_{\mu}}$. { Here, $k_{B}$ signifies the Boltzmann constant (set to one for convenience) and $T$ is the absolute temperature. In the matrix representation, Eq. \eqref{eq10} can be written as below
\begin{equation}
\hat{\varrho}_{th} = \frac{1}{Z} \left(
\begin{array}{cccc}
 \varrho_{11} & 0 & 0 & \varrho_{14} \\
 0 & \varrho_{22} & \varrho_{23} & 0 \\
 0 & \varrho_{32} & \varrho_{33} & 0 \\
 \varrho_{41} & 0 & 0 & \varrho_{44} \\
\end{array}
\right),
\label{eq88}
\end{equation}
and its nonzero elements read (setting $B=1$)}
\begin{align}
\small
    \varrho_{11} &= \frac{e^{(\mathcal{E} - \Delta) / T} [\mathcal{E} - \sqrt{2}\sin(\theta + \pi/4)]}{2 \mathcal{E}} \nonumber \\
    &\quad + \frac{e^{-(\Delta + \mathcal{E}) / T} [\mathcal{E} + \sqrt{2}\sin(\theta + \pi/4)]}{2 \mathcal{E}},\notag
\end{align}
\begin{equation}
\small
\varrho_{22} = e^{\Delta / T} \left(\frac{\sqrt{2}\sinh(\mathcal{R} / T) \sin(\theta - \pi/4)}{\mathcal{R}} + \cosh(\mathcal{R} / T)\right),\notag
\end{equation}
\begin{equation}
    \small
    \varrho_{33} = e^{\Delta / T} \left(\frac{\sqrt{2}\sinh(\mathcal{R} / T) \cos(\theta - \pi/4)}{\mathcal{R}} + \cosh(\mathcal{R} / T)\right),\notag
\end{equation}
\begin{align}
\small
    \varrho_{44} &= \frac{e^{(\mathcal{E} - \Delta) / T} [\mathcal{E} + \sqrt{2}\sin(\theta + \pi/4)]}{2 \mathcal{E}} \nonumber \\
    &\quad - \frac{e^{-(\Delta + \mathcal{E}) / T} [-\mathcal{E} + \sqrt{2}\sin(\theta + \pi/4)]}{2 \mathcal{E}},\notag
\end{align}
\begin{equation}
    \small
    \varrho_{14} = \varrho_{41}^* = \frac{2i \chi e^{-\Delta / T} \sinh(\mathcal{E} / T)}{\mathcal{E}},\notag
\end{equation}
\begin{equation}
\small
    \varrho_{23} = \varrho_{32}^* = -\frac{2i \phi^* e^{\Delta / T} \sinh(\mathcal{R} / T)}{\mathcal{R}},\notag
\end{equation}
with
\begin{align}
\small
    Z &= 2 \bigg\{\cosh \left(\frac{\Delta }{T}\right) \left[\cosh \left(\frac{\mathcal{E}}{T}\right) + \cosh \left(\frac{\mathcal{R}}{T}\right)\right] \nonumber \\
    &\quad + \sinh \left(\frac{\Delta }{T}\right) \left[\cosh \left(\frac{\mathcal{R}}{T}\right) - \cosh \left(\frac{\mathcal{E}}{T}\right)\right]\bigg\},
\end{align}
where $\mathcal{R}= \sqrt{4 D_{z}^2 - \sin(2 \theta) + 4 J^2 + 1}$, $\phi = D_{z} + i J$, $\mathcal{E} = \sqrt{1 + 4 G_{z}^2 + 4 J^2 \gamma^2 + \sin(2 \theta)}$, and $\chi = G_{z} + i \gamma J$. 

\subsection{Charging the QB}
{
The QB can be charged by applying a local external magnetic field either in the $x$-direction or the $y$-direction, with the charging field strength $\Omega = \Omega(t)$, which can be time-dependent or time-independent. For simplicity, we assume a time-independent charging field. The charging Hamiltonian is given by:
\begin{equation}
    \hat{\mathcal{H}}_{\mathcal{C}}^{(x,y)} = \Omega \left(\hat{\sigma}_{1}^{(x,y)} \otimes \hat{\mathbb{I}}_2 + \hat{\mathbb{I}}_2 \otimes \hat{\sigma}_{2}^{(x,y)}\right).
    \label{eq20}
\end{equation}

The magnetic field is applied uniformly to all qubits in the system. As the battery reaches its maximum stored energy at time $t = \tau$, the charging field must be disconnected to prevent the battery from reverting back to an uncharged state due to continuous cycling.

Both the Pauli-$X$ and Pauli-$Y$ gates can equivalently serve as charging operators for the QB. The Pauli-$X$ gate acts as a bit-flip, while the Pauli-$Y$ gate performs both bit-flip and phase-flip operations. One can use either Pauli-$X$ or Pauli-$Y$ gates for charging because both act as unitary operations that drive the QB system from its initial state to a charged state. The Pauli-$X$ gate is a pure bit-flip operator, while the Pauli-$Y$ gate incorporates both bit-flip and phase-flip, providing more versatility depending on the QB's configuration and field interactions.

In a closed system, the charging process is implemented as a unitary evolution:
\begin{equation}
    \hat{U}_{\mathcal{C}} (t) = \exp(-i \hat{\mathcal{H}}_{\mathcal{C}}t),
    \label{eq:unitary_operator}
\end{equation}
which governs the cyclic charging of the QB over time.

For the Pauli-$Y$ gate charging Hamiltonian, the unitary operator takes the form:
\begin{equation}
    \hat{\mathcal{U}}_{\mathcal{C}}^{(y)}(t) = \left(
    \begin{array}{cccc}
       \alpha & \lambda & \lambda & \beta \\
        \lambda & \alpha & \beta & \lambda \\
        \lambda & \beta & \alpha & \lambda \\
        \beta & \lambda & \lambda & \alpha \\
    \end{array}
    \right),
    \label{eq:unitary_y_gate}
\end{equation}
where $\alpha = \cos^2(\Omega t)$, $\beta = -\sin^2(\Omega t)$, and $\lambda = -\frac{i}{2}\sin(2 \Omega t)$.

Similarly, for the Pauli-$X$ gate charging Hamiltonian, the unitary operator is:
\begin{equation}
    \hat{U}_{\mathcal{C}}^{(x)} (t) = \left(
    \begin{array}{cccc}
        \alpha & \vartheta_- & \vartheta_- & 1-\alpha \\
        \vartheta_+ & \alpha & \alpha-1 & \vartheta_- \\
        \vartheta_+ & \alpha-1 & \alpha & \vartheta_- \\
        1-\alpha & \vartheta_+ & \vartheta_+ & \alpha \\
    \end{array}
    \right),
    \label{eq:unitary_x_gate}
\end{equation}
where $\vartheta_{\pm} = \pm \frac{1}{2} \sin(2 \Omega t)$.

During the charging process, the evolution of the system is determined solely by the charging Hamiltonian $\hat{\mathcal{H}}_{\mathcal{C}}^{(x,y)}$, while the Hamiltonian $\hat{\mathcal{H}}_{\mathcal{Q}}$ is relevant before and after charging.
}
\subsection{Performance indicators for QBs}
Let $\hat{\varrho}$ be a quantum state of QB with Hamiltonian $\hat{\mathcal{H}}_{\mathcal{Q}}$. The question we are interested in is: how much work can be extracted from this QB through a cyclic unitary process? In this context, cyclicity implies that the Hamiltonian of the QB must be the same at the start and end of the process, i.e., $\hat{\mathcal{H}}_{\mathcal{Q}} = \hat{\mathcal{H}}_{\mathcal{Q}}(0) = \hat{\mathcal{H}}_{\mathcal{Q}}(\tau)$ \cite{ binder2015quantum}. Since the evolution is unitary, any change in the internal energy $\langle \hat{\mathcal{H}}_{\mathcal{Q}} \rangle$ is attributed to the work done on or by QB.

To analyze the QB's Hamiltonian, we express it in terms of its increasing spectral decomposition, namely
\begin{equation}
\hat{\mathcal{H}}_{\mathcal{Q}} := \sum_{\mu=1}^{4} \nu_\mu |\psi_\mu\rangle\langle\psi_\mu|, \text{ where } \nu_{\mu+1} \geq \nu_\mu \; \forall \mu.
\end{equation}
In this expression, $\nu_\mu$ denotes the energy eigenvalues, and $|\psi_\mu\rangle$ represents the corresponding eigenstates. This formulation enables the evaluation of energy variations and potential for work extraction in a cyclic unitary process, where the QB's Hamiltonian remains unchanged at the start and end of the cycle.

The state $\hat{\varrho}$, on the other hand, is expressed in its decreasing eigen-decomposition

\begin{equation}
\hat{\varrho} := \sum_{\kappa=1}^{4} r_{\kappa} |r_\kappa\rangle\langle r_\kappa|, \text{ with } r_{\kappa+1} \leq r_\kappa \; \forall \kappa.
\end{equation}

The goal is to transform $\hat{\varrho}$ into a state with lower internal energy, extracting the difference in internal energy in the process.

After maximal cyclic, unitary work extraction, no further work can be extracted, and the system ends up in a so-called passive state $\hat{\pi}$ \cite{ghosh2020enhancement, binder2015quantum, pusz1978passive, allahverdyan2004maximal}. A passive state is unique up to degeneracy in the Hamiltonian. Passive states are diagonal in the Hamiltonian's eigenbasis, with decreasing populations for increasing energy levels \cite{allahverdyan2004maximal, binder2015quantum}. That is, a state $\hat{\varrho}$, as defined above, is passive if $|r_n\rangle = |\psi_n \rangle \; \forall n$. Gibbs thermal states are consequently passive \cite{binder2015quantum}; therefore, we set $\hat{\pi}=\hat{\varrho}_{th}$.

{ The maximum work that can be extracted from a non-passive state $\hat{\varrho}$ for a Hamiltonian $\hat{\mathcal{H}}_{\mathcal{Q}}$ via a cyclic unitary process $\hat{\varrho}_{th} \to \hat{\varrho}$ ($\hat{\varrho}=\hat{U}_{\mathcal{C}} \hat{\varrho}_{th}\hat{U}_{\mathcal{C}}^\dagger$) \cite{ghosh2020enhancement}, called ergotropy, is found to be \cite{carrasco2110collective,allahverdyan2004maximal}

\begin{equation}
\xi(t) :=\sum_{m,n} r_m \nu_n \left[|\langle \psi_n | r_m \rangle|^2 - \delta_{mn}\right], 
\end{equation}
where $\delta_{mn}$ is the Kronecker delta function.  Equivalently, one can write the above expression as \cite{zakavati2021bounds}
\begin{equation}
\xi(t) := \text{Tr}[(\hat{\varrho} - \hat{\varrho}_{th}) \hat{\mathcal{H}}_{\mathcal{Q}}].
\end{equation}
If the final state is another non-passive state with relatively higher energy than $\hat{\pi}$, say $\hat{\Pi}$, then work done by the QB can be defined as \cite{ghosh2020enhancement}

\begin{equation}
\mathcal{W}(t)=\text{Tr}[(\hat{\varrho} -\hat{\Pi})\mathcal{\hat{H}}_{\mathcal{Q}}].
\end{equation}

Notice that the average power is defined as $\langle p(t)\rangle=\mathcal{W}(t)/t$, and the peak average power is $\langle p_{\max}(t)\rangle= \max\limits_{t} \mathcal{W}(t)/t$ \cite{ghosh2020enhancement}.}
Besides, the efficiency $\eta = \mathcal{W}(t)/\xi(t)$ quantifies the fraction of stored energy $\mathcal{W}(t)$ that can be extracted as ergotropy $\xi(t)$. Efficiency is typically constrained to $\eta \leq 1$, with $\eta = 1$ indicating $\mathcal{W}(t)=\xi(t)$ for which the initial state should be a passive state.

\section{Results and discussion}\label{sec:3}
The study of temporal ergotropy evolution in various Heisenberg model subclasses, detailed in Section \ref{sec:2}, provides key insights into the charging and discharging dynamics of spin-based QB systems, facilitating their output optimization. { The closed-form expression for ergotropy based on our considered model with $Y$-gate charging Hamiltonian is given by}

\begin{widetext}
\begin{equation}\label{ergotropy17}
\small
\begin{split}
\xi(t) = & \frac{2 \sin ^2(\Omega t)}{\sqrt{ab} \left(2 h \epsilon _a+d^2+1\right)}  \bigg\{c \left[-2 \sqrt{ab} h \epsilon _a+\sqrt{a} \cos (2 \Omega t) \left(-2 \sqrt{b} d g \epsilon _a+\sqrt{b} \left(d^2+1\right)+2 \gamma  \left(d^2-1\right) J\right)+2 \sqrt{ab} d \epsilon _b\right]  \\
& + 2 \sqrt{b} f_a \big[h (a+2 J (c-2 J))+2 c d g J \cos (2 \text{$\Omega $t})+\sin (2 \theta ) (h-d g)\big]+2 \sqrt{a} d f_b \big[b+2 \gamma  J (c-2 \gamma  J)\big]\bigg\},
\end{split}
\end{equation}
where
\( a = \mathcal{R}^2 \),
\( b = \mathcal{E}^2 \),
\( c = -\Delta +\gamma  J+J \),
\( d = e^{\frac{\sqrt{b}}{T}} \),
\( \epsilon _a = \cosh \left(\frac{\sqrt{a}}{T}\right) \), \( \epsilon _b = \cosh \left(\frac{\sqrt{b}}{T}\right) \),
\( f_a = \sinh \left(\frac{\sqrt{a}}{T}\right) \), \( f_b = \sinh \left(\frac{\sqrt{b}}{T}\right) \),
\( g = e^{\frac{2 \Delta}{T}} \), and
\( h = e^{\frac{2 \Delta + \sqrt{b}}{T}} \).
\end{widetext}

\subsection{Effect of variable Zeeman { splitting}}
We investigate the temporal evolution of ergotropy \eqref{ergotropy17} under an inhomogeneous magnetic field parameterized by $\theta$ in Eq. \eqref{eq6}, focusing on QB performance over time $\Omega t$ and $\theta$, with DM and KSEA couplings set to zero for various Heisenberg model subclasses. Setting $B=1$ and $0 \leq \theta \leq \frac{\pi}{2}$, Zeeman splitting affects one spin at $\theta=0$ or $\theta=\frac{\pi}{2}$, while both spins experience equal splitting at $\theta=\frac{\pi}{4}$. We constrain $\Omega t$ to $0 \leq \Omega t \leq 2\pi$, allowing the QB to undergo multiple charging and discharging cycles at $T = 0.1$, illustrating cyclic unitary charging operation. {  A common strategy, often adopted in experimental protocols, involves deactivating the NOT-gate-based charging field once maximum ergotropy is reached. This prevents the quantum NOT-gate-based charging Hamiltonian, as described in Eq.~\eqref{eq20}, from returning the system to its thermal state. Without deactivation, the system would oscillate between charged and passive states over time. It is important to note that this approach is well-established and not a novel execution of this work.}

\begin{figure}[h]
    \centering
\includegraphics[width=\columnwidth]{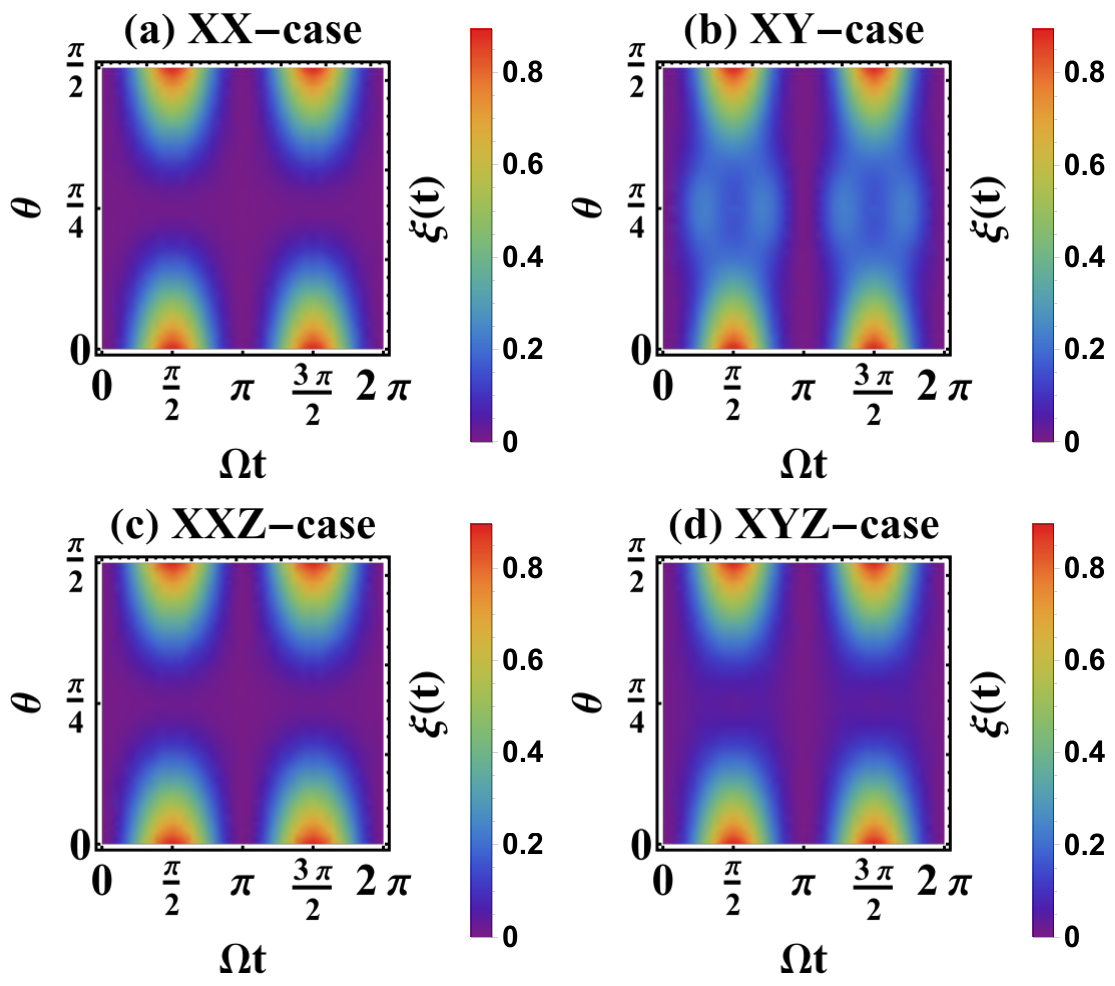}
    \caption{ Density plots of $\xi(t)$ in the AFM scenario versus $\Omega t$ and $\theta$, with $D_z = 0$, $G_z = 0$, $T = 0.1$, and $J = 1$: (a) $XX$ ($\Delta = 0$, $\gamma = 0$); (b) $XY$ ($\Delta = 0$, $\gamma = 0.5$); (c) $XXZ$ ($\Delta = 0.5$, $\gamma = 0$); (d) $XYZ$ ($\Delta = 0.5$, $\gamma = 0.5$).}
    \label{f2}
    \end{figure}

The density plots in Fig. \ref{f2} show the variation of ergotropy $\xi(t)$ in the { AFM scenario ($J=1$), with $\gamma\geq0$, $\Delta\geq0$, $D_z=0$, $G_z=0$, and $T=0.1$, as functions of time $\Omega t$ and Zeeman splitting inhomogeneity parameter $\theta$.} Maximum $\xi(t\rightarrow \tau)$ occurs when the magnetic field affects one spin at $\theta=0$ or $\theta=\frac{\pi}{2}$, while the lowest $\xi(t)$ is observed when both spins are equally affected at $\theta=\frac{\pi}{4}$. Although trends { appear similar} across HSC cases, the $XY$ model exhibits distinct behavior around $\theta=\frac{\pi}{2}$, maintaining nonzero $\xi(t)$ at $\theta=\frac{\pi}{4}$ { and suggesting slightly better extractable work compared to the $XX$, $XXZ$, and $XYZ$ models.}
    \begin{figure}[h]
    \centering    \includegraphics[width=\columnwidth]{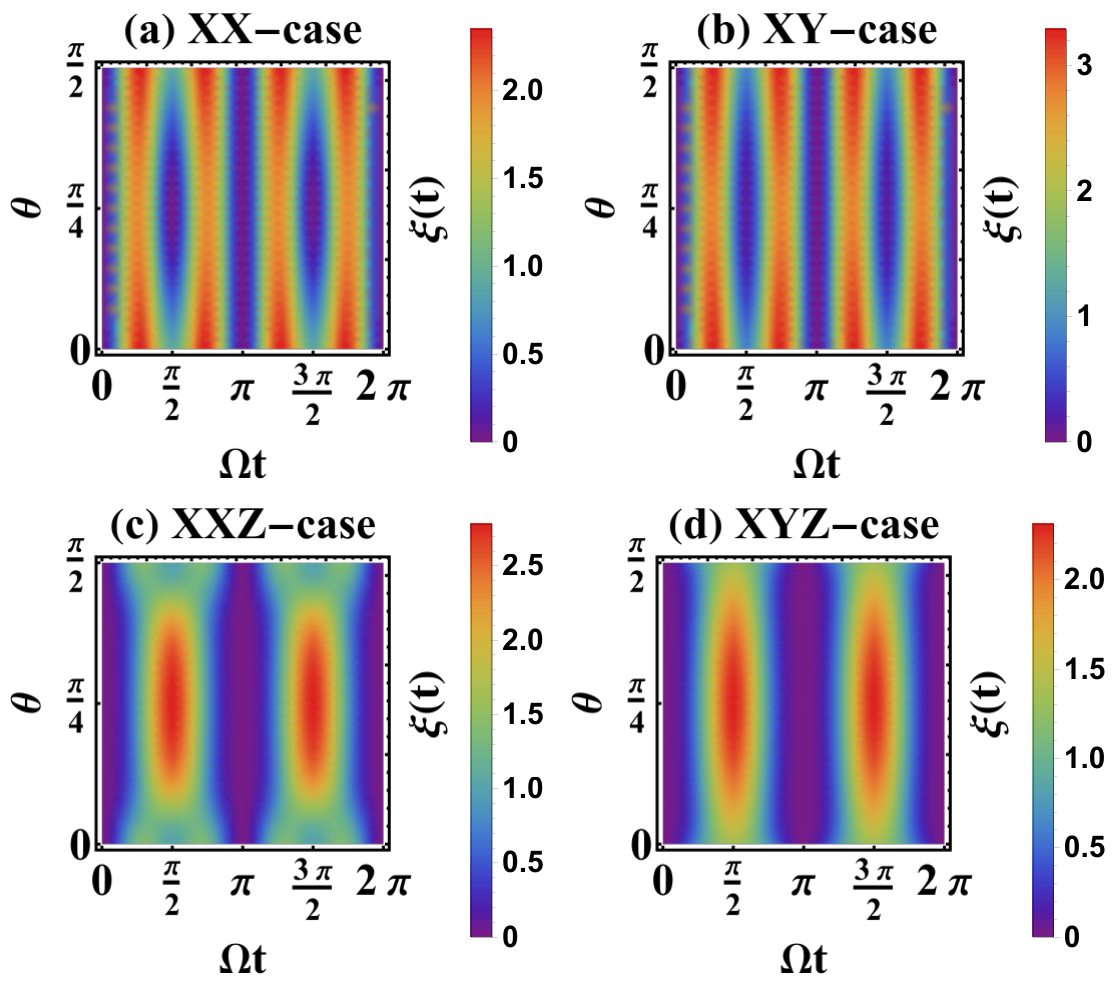}
      \caption{Density plots of $\xi(t)$ in the FM scenario as functions of $\Omega t$ and $\theta$, with $D_z = 0$, $G_z = 0$, $T = 0.1$, and $J = -1$: (a) $XX$ ($\Delta = 0$, $\gamma = 0$); (b) $XY$ ($\Delta = 0$, $\gamma = 0.5$); (c) $XXZ$ ($\Delta = -0.5$, $\gamma = 0$); (d) $XYZ$ ($\Delta = -0.5$, $\gamma = 0.5$)}
    \label{f3}
\end{figure}
Within the interval $0 \leq \Omega t \leq 2\pi$, the QB experiences two complete charge-discharge cycles. The maximal charge states are denoted by red regions along the $\Omega t$-axis. For the given set of parameters, the optimal charging time is determined to be $\tau = \frac{\pi}{2\Omega}$. The peak value of $\xi_{max}(t)$ remains invariant in time due to the unitary nature of the charging process.

We have analyzed the effect of inhomogeneous Zeeman splitting on the QB in the AFM case over time. Next, we will investigate the FM case. { The density plots in Fig.~\ref{f3} show the temporal evolution of $\xi(t)$  as functions of $\Omega t$ and $\theta$ in the FM scenario,  where $J=-1$ and $\Delta\leq0$ with $\gamma\geq0$, $D_z=0$, $G_z=0$, and $T=0.1$. Obviously, the results in the FM case differ significantly from those in the AFM case.}

Unlike the AFM case illustrated in Fig. \ref{f2}, where the highest values of $\xi(t\rightarrow \tau)$ are attained at $\tau=\frac{\pi}{2\Omega}$ for all considered HSC cases and at $\theta=0$ or $\frac{\pi}{2}$, the FM scenario presents a distinct behavior. In the $XX$ model (Fig. \ref{f3}(a)) and the $XY$ model (Fig. \ref{f3}(b)), the peak value of ergotropy, or first full-charging, occurs at $\tau=\frac{\pi}{4\Omega }$. This peak is reached four times within the interval $0 \leq \Omega t \leq 2\pi$. { This shows that, in the FM case, the charging rate for the $XX$ and $XY$ models in Figs. \ref{f3}(a)-(b) is two times faster than in the AFM scenario for all the HSC cases depicted in Figs. \ref{f2}(a)-(d). For the AFM case, the peak value of ergotropy is achieved for the Zeeman splitting parameter at $\theta=0$ (or equivalently $\frac{\pi}{2}$), whereas in the FM scenario, the ergotropy value remains nonzero for all values of $0 \leq \theta \leq \frac{\pi}{2}$, with maxima at $\theta=0$ and $\theta=\frac{\pi}{2}$.} Thus, the inhomogenous Zeeman splitting in the { FM case with $XX$ and $XY$ models} does not substantially alter the maximum ergotropy.

{ Considering $XXZ$ and $XYZ$ models from Figs. \ref{f3}(c) and \ref{f3}(d), one can see that unlike Figs. \ref{f3}(a) and \ref{f3}(b) where the impact of Zeeman splitting parameter $\theta$ is roughly independent on ergotropy, here we find that the maximum value of $\xi(t\rightarrow \tau)$ occurs around $\theta=\frac{\pi}{4}$ wheres minimum value but not zero occurs at $\theta=0$ or equivalently $\frac{\pi}{2}$. Furthermore, in contrast to Figs. \ref{f3}(a) and \ref{f3}(b), where the first maximum charge is achieved at $t = \pi/4\Omega$, the first peak of $\xi(t)$ in Figs. \ref{f3}(c) and \ref{f3}(d) occurs at $t = \pi/2\Omega$.}

A key observation is that the overall peak ergotropy in all HSC categories under the FM scenario is nearly three times greater than the peak ergotropy in the AFM case. { Notably, the Heisenberg $XY$ model demonstrates greater robustness compared to the other three models for QB in both FM and AFM scenarios.

When optimizing a QB with HSC models, the strategies differ} significantly between FM and AFM scenarios. For the FM case, where spin polarizations are parallel, the $XX$ and $XY$ models achieve the fastest charging rates, with peak ergotropy occurring at $\tau = \frac{\pi}{4\Omega}$ and repeating four times within $0 \leq \Omega t \leq 2\pi$. In contrast, the $XXZ$ and $XYZ$ models reach maximum ergotropy at $\theta = \frac{\pi}{4}$ and have slower charging, peaking at $\tau = \frac{\pi}{2\Omega}$. This indicates efficient energy transfer in the FM scenarios, but the choice of coupling model affects the timing of optimal charging.
{ For AFM scenario, where spins are oppositely aligned, maximum ergotropy for all models occurs at $\theta = 0$ or $\frac{\pi}{2}$. The $XY$ model shows minimal ergotropy around $\theta = \frac{\pi}{4}$, suggesting that the AFM setting benefits from specific spin chain couplings to enhance performance. Thus, according to obtained results from Figs. \ref{f2} and \ref{f3}, we will use $\theta = \frac{\pi}{4}$ for the FM case and $\theta = 0$ or $\frac{\pi}{2}$ for the AFM case in the continuation of our analysis.}
\subsection{Effect of Temperature}
\begin{figure}[t]
    \centering    \includegraphics[width=\columnwidth]{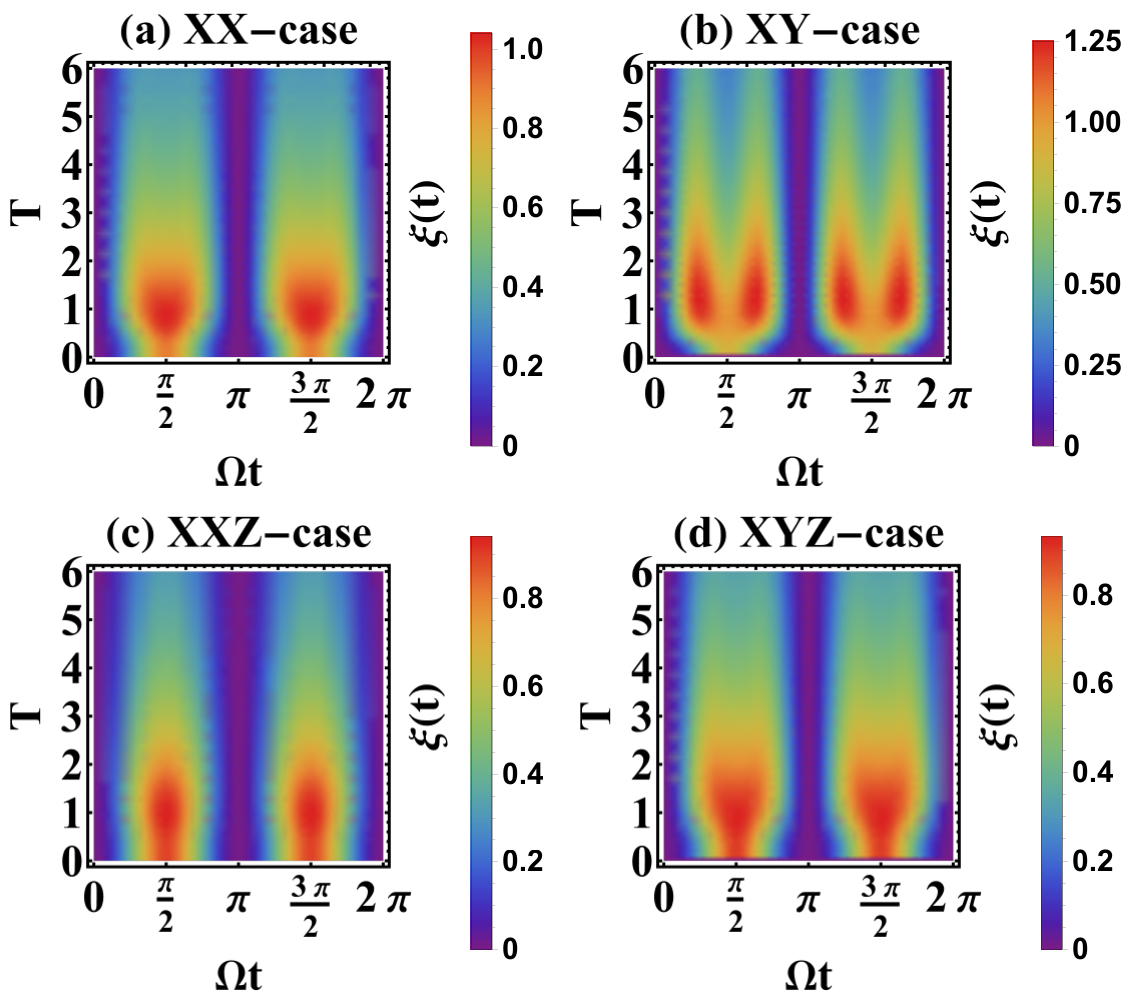}
        \caption{Density plots of $\xi(t)$ in the AFM scenario versus $\Omega t$ and $T$, with $D_z = 0$, $G_z = 0$, $\theta = 0$ (or $\frac{\pi}{2}$), and $J = 1$: (a) $XX$ ($\Delta = 0$, $\gamma = 0$); (b) $XY$ ($\Delta = 0$, $\gamma = 0.5$); (c) $XXZ$ ($\Delta = 0.5$, $\gamma = 0$); (d) $XYZ$ ($\Delta = 0.5$, $\gamma = 0.5$).}
    \label{f4}
\end{figure}
Let's study how temperature $T$ affects $\xi(t)$ without DM and KSEA couplings in FM and AFM cases.

{
In the AFM scenario depicted in Fig. \ref{f4}, the behavior of $\xi(t)$ as functions of temperature $T$ and $\Omega t$ exhibits a nonmonotonic pattern across all subcategories of the HSC models. For instance, in Fig. \ref{f4}(a), $\xi(t)$ starts from a nonzero value at $T=0$, rises as $T$ increases, peaks around $T \approx 1$, and then gradually decreases as $T$ continues to rise, eventually approaching zero at asymptotically high temperatures. A similar trend is observed in Figs. \ref{f4}(c) and \ref{f4}(d), where the nonmonotonic dependence of $\xi(t)$ on $T$ is also evident. However, in Fig. \ref{f4}(b), the highly charged regions split from a single nonzero ergotropy region, depicted by the yellowish-red area, into two distinct red regions as $T$ increases. Despite this bifurcation, the overall nonmonotonic effect of $T$ on $\xi(t)$ persists. This bifurcation behavior also indicates a faster charging rate in the case illustrated in Fig. \ref{f4}(b). As previously discussed, in the $XY$ model, $\xi(t)$ does not peak at $T=0$ but instead starts from a lower value and increases with temperature, reaching its maximum around $T \approx 1$. This signifies a higher charging rate compared to the $XX$, $XXZ$, and $XYZ$ models, where the influence of temperature is less pronounced. In particular, the charging rate in the $XY$ model is nearly double that of the other three models for given fixed values of parameters. Moreover, the peak value of ergotropy is largest in the $XY$ case ($\xi(t)\approx 1.25)$, followed by $XX$ ($\xi(t)\approx 1)$, with $XXZ$ and $XYZ$ having comparatively lower peak values ($\xi(t)\approx 0.9)$.
}

\begin{figure}[h]
    \centering    \includegraphics[width=\columnwidth]{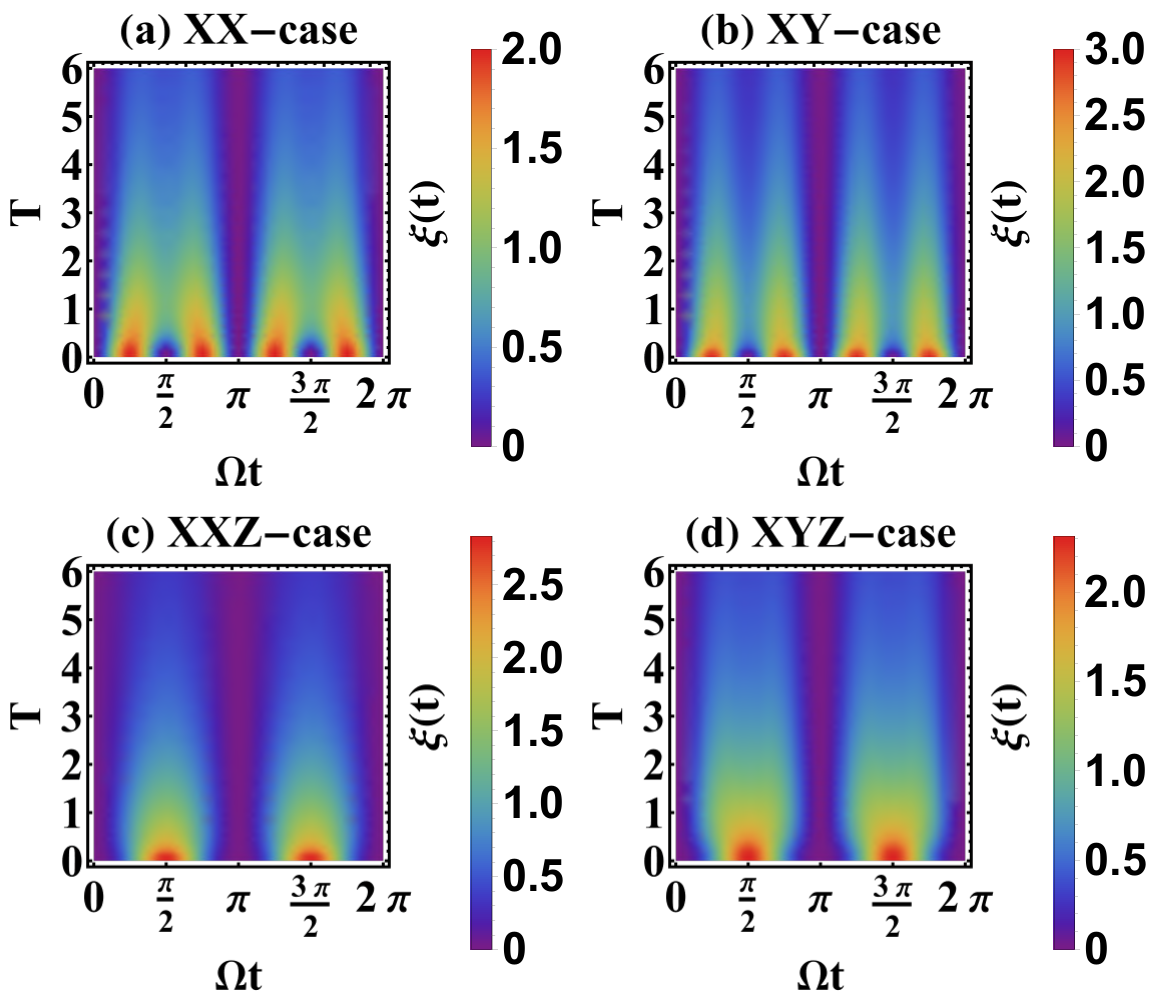}
\caption{Density plots of $\xi(t)$ in the FM scenario versus $\Omega t$ and $T$, with $D_z = 0$, $G_z = 0$, $\theta = \frac{\pi}{4}$, and $J = -1$: (a) $XX$ ($\Delta = 0$, $\gamma = 0$); (b) $XY$ ($\Delta = 0$, $\gamma = 0.5$); (c) $XXZ$ ($\Delta = -0.5$, $\gamma = 0$); (d) $XYZ$ ($\Delta = -0.5$, $\gamma = 0.5$).}
\label{f5}
\end{figure}

{ In contrast to the AFM scenario depicted in Fig. \ref{f4}, the FM case illustrated in Fig. \ref{f5} reveals a significant decrease in $\xi(t)$ with increasing $T$, displaying a monotonic behavior. Unlike the AFM case, where the variation of $\xi(t)$ with $T$ is nonmonotonic, the FM scenario shows a steady decline as $T$ rises. This monotonic decrease in $\xi(t)$ is attributed to the more fragile nature of quantum coherence and correlations in FM configurations of the HSC models, which are prone to degradation with increasing $T$ compared to the stronger AFM scenario. This fragility is consistently reflected in the $\xi(t)$ plots across all $T$.

Furthermore, the charging rates differ significantly between the FM and AFM cases. In the FM scenario, the peak charging rates for the $XX$ and $XY$ models are higher than those for the $XXZ$ and $XYZ$ models. The $XY$ model exhibits the highest $\xi(t)$, followed by the $XX$ model, with the $XXZ$ model showing lower peak of $\xi(t)$, and the $XYZ$ model having the lowest. This is in contrast to the AFM case, where the nonmonotonic behavior allows for temperature-dependent variations that affect the peak values differently. In both scenarios, however, the $XY$ model consistently demonstrates the highest peak of $\xi(t)$, highlighting its superior performance in both FM and AFM cases.

Overall, temperature has a more detrimental impact on ergotropy in the FM case across all models  compared to the AFM scenario, indicating reduced performance of QBs as temperature increases in the FM configuration. This suggests that QBs operating under FM conditions are more sensitive to thermal effects. For optimal performance, the strategy should focus on maintaining low temperatures and leveraging the specific strengths of each model, whether it be maximizing charging rates or achieving higher peak ergotropy. In particular, models like $XY$ consistently show better ergotropy production, both in FM and AFM cases, making them more suitable for efficient QB performance for given fixed values of other parameters.
}
\subsection{Symmetric-antisymmetric exchange coupling}
\subsubsection{Effect of KSEA Coupling}

In addition to the fact that a high ergotropy in both AFM and FM cases requires low temperatures, it is crucial to apply Zeeman splitting fields effectively. { As can be seen from the results, applying the magnetic field to one spin is advantageous in AFM scenarios, while in FM context, applying it to both spins is beneficial.} Hence, we investigate how DM and KSEA couplings can be utilized to enhance ergotropy in HSC-based QBs using predefined parameters from previous analyses.

\begin{figure}[t]
    \centering
    \includegraphics[width=\columnwidth]{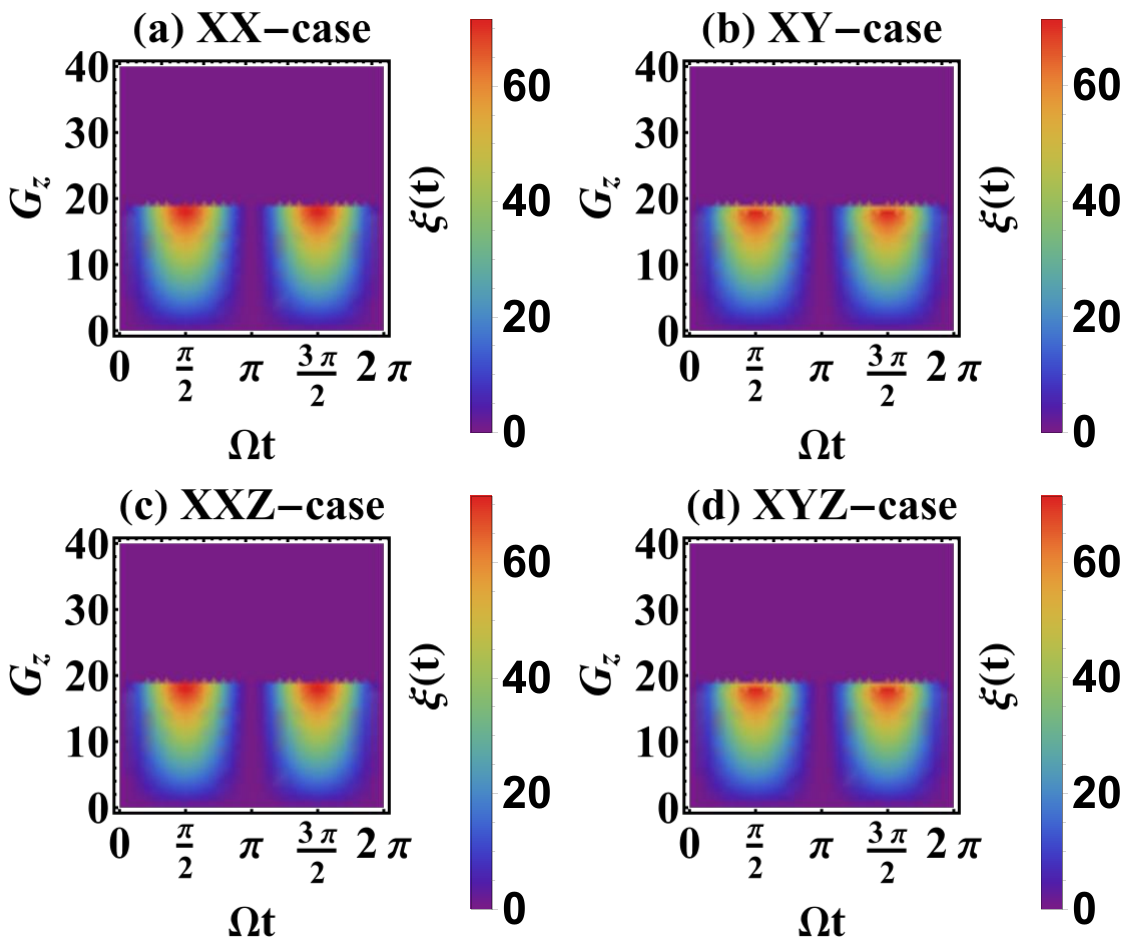}
     \caption{Density plots of $\xi(t)$ in the AFM scenario versus $\Omega t$ and $G_z$, with $D_z = 0$, $T = 0.1$,  $\theta = 0$ (or $\frac{\pi}{2}$), and $J = 1$: (a) $XX$ ($\Delta = 0$, $\gamma = 0$); (b) $XY$ ($\Delta = 0$, $\gamma = 0.5$); (c) $XXZ$ ($\Delta = 0.5$, $\gamma = 0$); (d) $XYZ$ ($\Delta = 0.5$, $\gamma = 0.5$).}
     \label{f6}
\end{figure}

Figure \ref{f6} shows the density plots of $\xi(t)$ in the AFM-case as functions of $\Omega t$ and the $z$-component of KSEA coupling at $T=0.1$, $\theta=\frac{\pi}{2}$, and $D_z=0$. It shows that $G_z$ positively affects all HSC cases. The full-charging occurs around $\tau=\frac{\pi}{2\Omega}$ during the first charge. Notably, $G_z$ significantly boosts ergotropy in the AFM case with peak values around 75 near $G_z\approx 20$, as seen in Fig. \ref{f6}. Beyond $G_z\approx 20$, ergotropy sharply drops to zero for $G_z>20$, indicating that $G_z$ has maximized the QB's charge, and further increases in $G_z$ do not enhance ergotropy. A similar trend is observed in the FM case, which is omitted here for brevity.

\begin{figure}[t]
    \centering
    \includegraphics[width=\columnwidth]{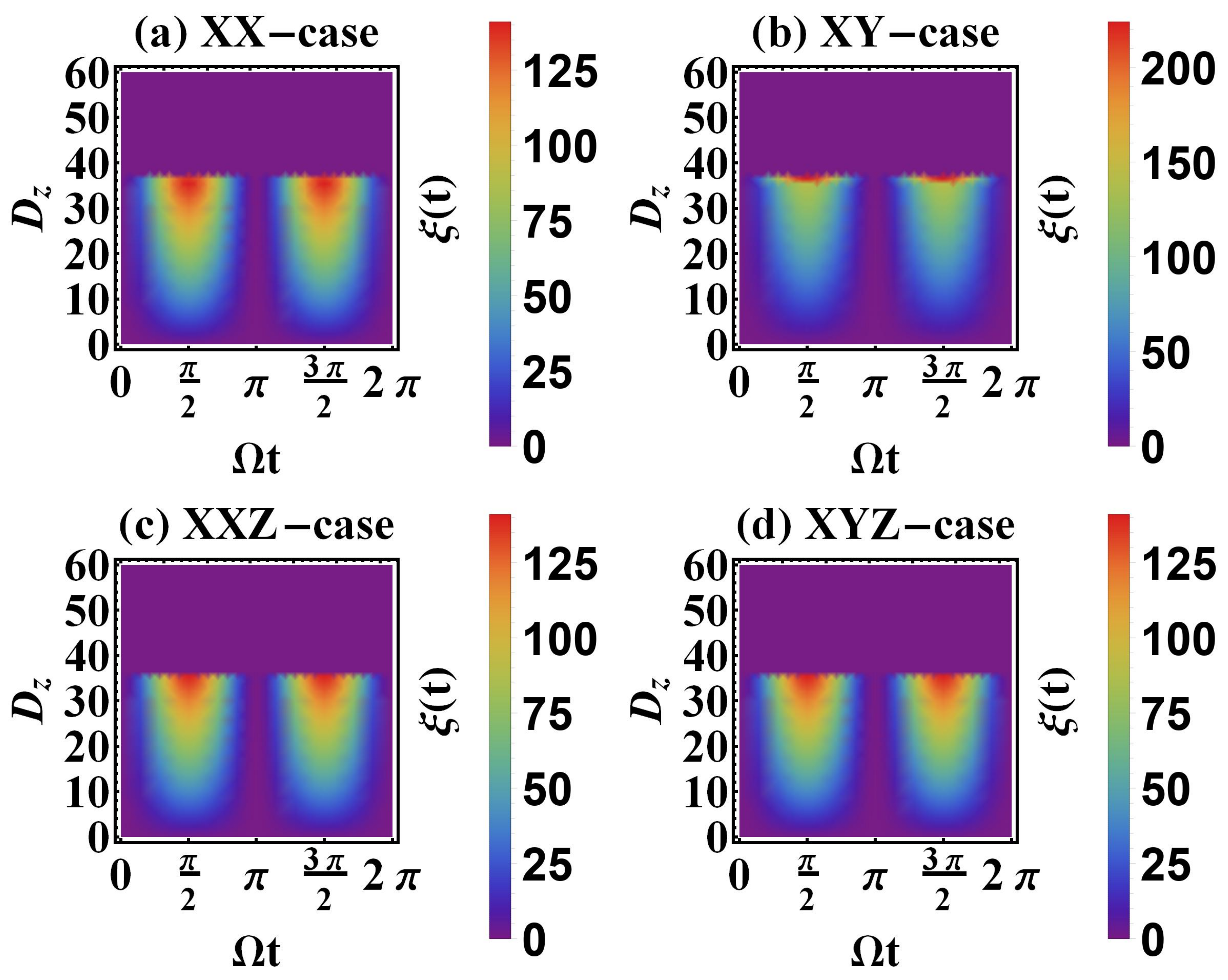}
          \caption{ Density plots of $\xi(t)$ in the AFM scenario versus $\Omega t$ and $D_z$, with $G_z = 0$, $T = 0.1$, $\theta = 0$ (or $\frac{\pi}{2}$), and $J = 1$: (a) $XX$ ($\Delta = 0$, $\gamma = 0$); (b) $XY$ ($\Delta = 0$, $\gamma = 0.5$); (c) $XXZ$ ($\Delta = 0.5$, $\gamma = 0$); (d) $XYZ$ ($\Delta = 0.5$, $\gamma = 0.5$).}
    \label{f7}
\end{figure}

\begin{figure}[t]
    \centering    \includegraphics[width=\columnwidth]{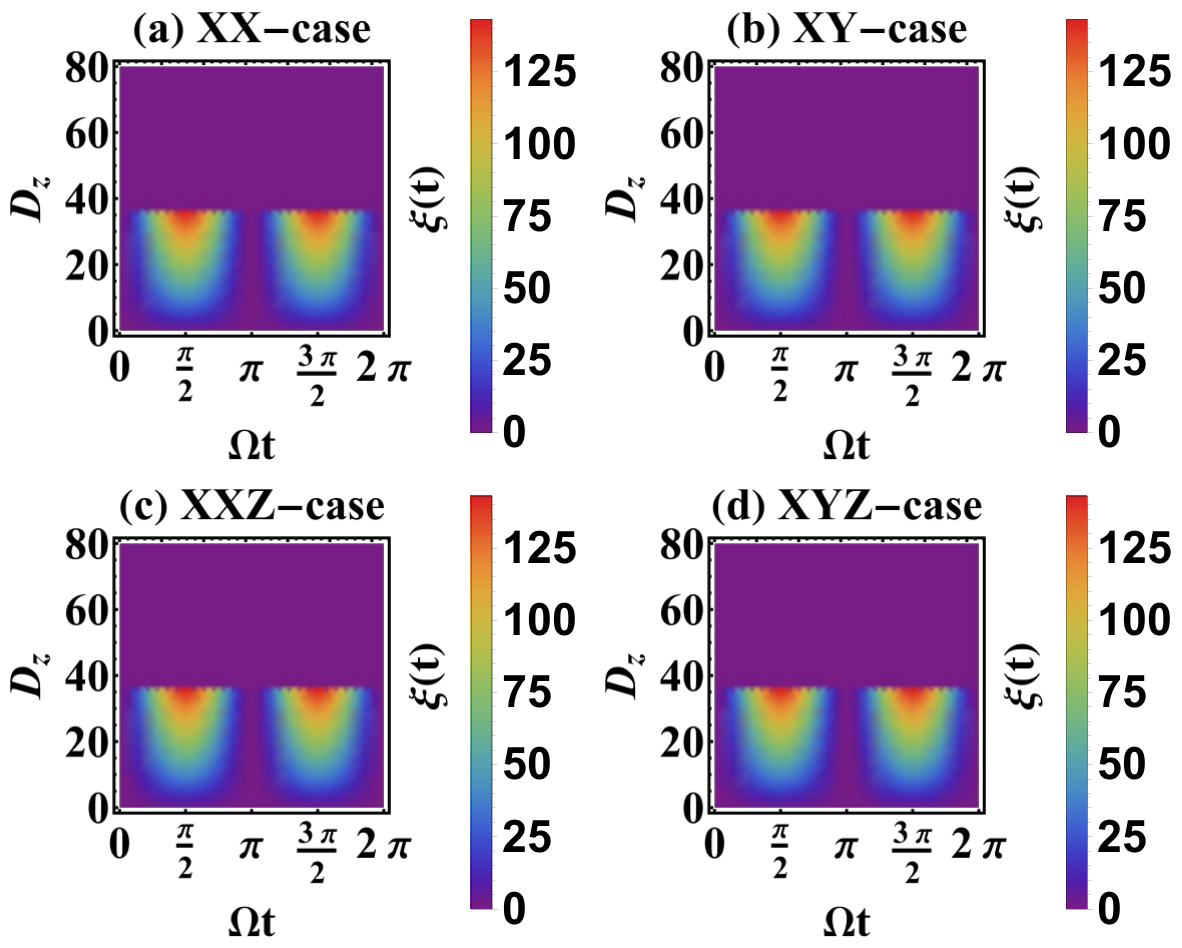}
         \caption{Density plots of $\xi(t)$ in the FM scenario versus $\Omega t$ and $D_z$, with $G_z = 0$, $T = 0.1$, $\theta = \frac{\pi}{4}$, and $J = -1$: (a) $XX$ ($\Delta = 0$, $\gamma = 0$); (b) $XY$ ($\Delta = 0$, $\gamma = 0.5$); (c) $XXZ$ ($\Delta = -0.5$, $\gamma = 0$); (d) $XYZ$ ($\Delta = -0.5$, $\gamma = 0.5$).}
    \label{f8}
\end{figure}

\begin{figure*}[!t]
\centering
\includegraphics[width=0.32\textwidth]{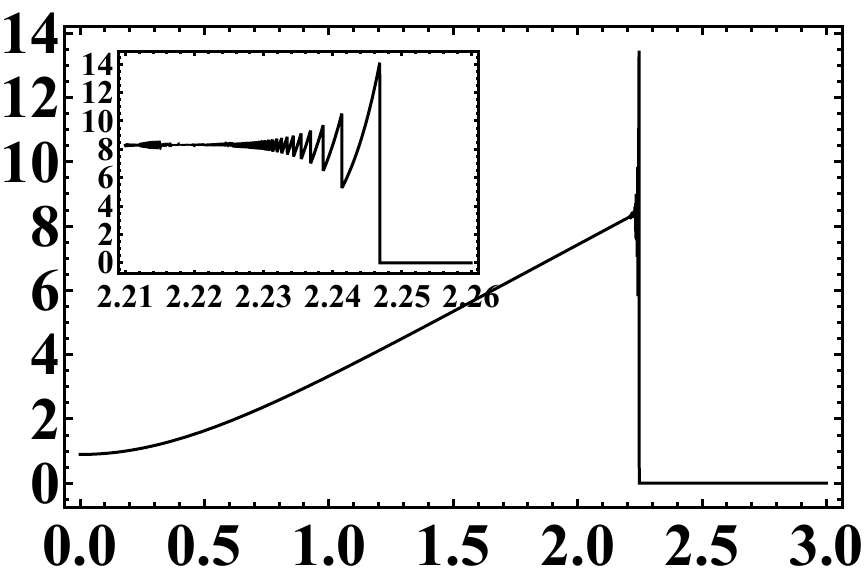}
\includegraphics[width=0.325\textwidth]{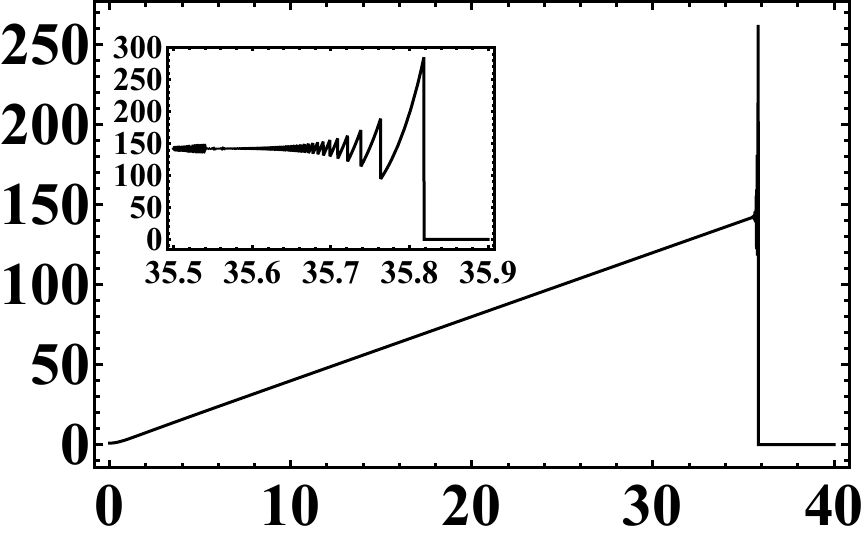}
\includegraphics[width=0.335\textwidth]{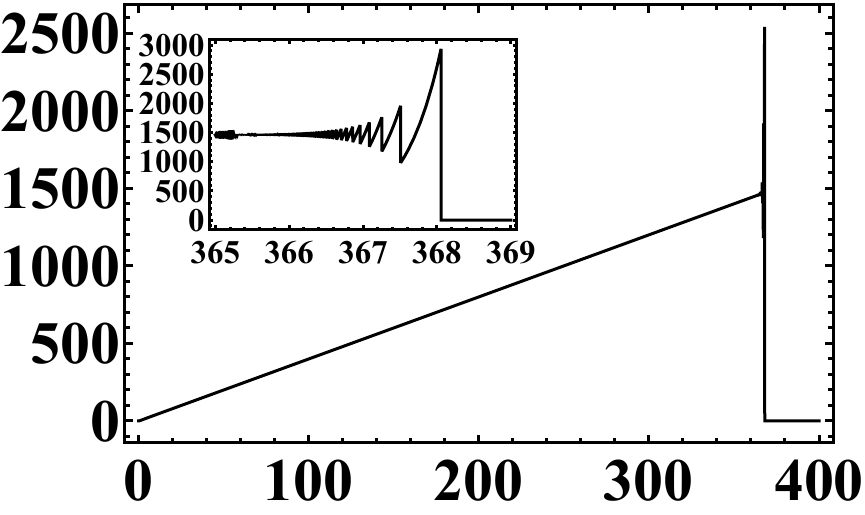}
\put(-515,45){\rotatebox{90}{$\boldsymbol{\xi_{\max}}$}}
\put(-362,90){$(\boldsymbol{a})$}
\put(-190,90){$(\boldsymbol{b})$}
\put(-20,90){$(\boldsymbol{c})$}
\put(-257,-8){$\boldsymbol{D_z}$}
\put(-93,-8){$\boldsymbol{D_z}$}
\put(-420,-8){$\boldsymbol{D_z}$}
\put(-260,25){$\boldsymbol{T=0.1}$}
\put(-95,25){$\boldsymbol{T=1.0}$}
\put(-450,25){$\boldsymbol{T=0.01}$}
\caption{Maximum ergotropy $\xi_{\text{max}}$ versus $D_z$ for the AFM case in the Heisenberg $XYZ$ model with parameters $G_z=0$, $J=1$, $\Delta=0.5$, $\gamma=0.5$, $\Omega=1$, and $\theta=\frac{\pi}{2}$. The subfigures represent different temperatures: (a) $T=0.01$, (b) $T=0.1$, and (c) $T=1$.}
\label{f9}
\end{figure*}

\begin{figure*}[!t]
\centering
\includegraphics[width=0.32\textwidth]{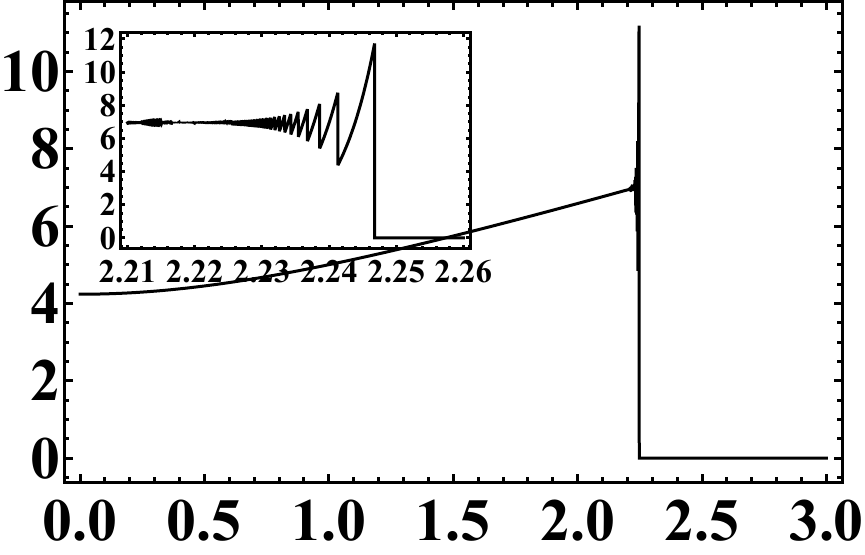}
\includegraphics[width=0.325\textwidth]{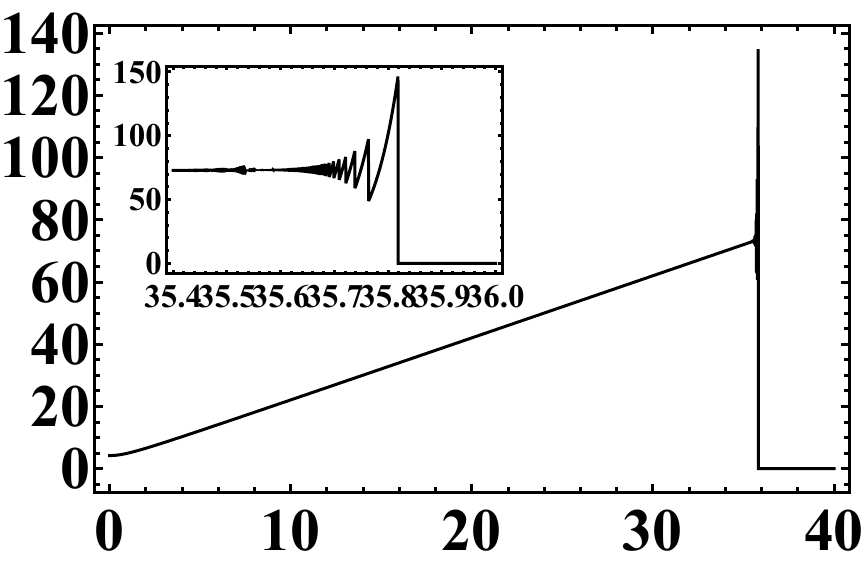}
\includegraphics[width=0.335\textwidth]{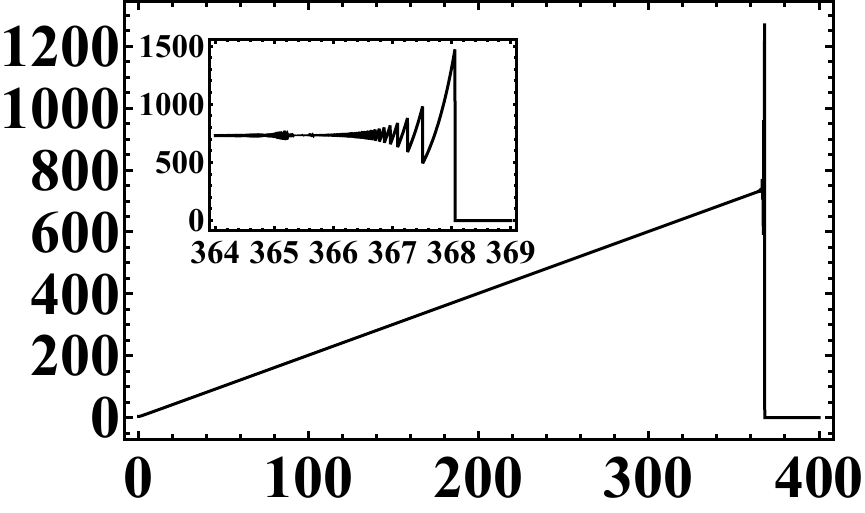}
\put(-510,52){\rotatebox{90}{$\boldsymbol{K}$}}
\put(-362,90){$(\boldsymbol{a})$}
\put(-190,90){$(\boldsymbol{b})$}
\put(-20,90){$(\boldsymbol{c})$}
\put(-257,-8){$\boldsymbol{D_z}$}
\put(-93,-8){$\boldsymbol{D_z}$}
\put(-420,-8){$\boldsymbol{D_z}$}
\put(-260,25){$\boldsymbol{T=0.1}$}
\put(-95,25){$\boldsymbol{T=1.0}$}
\put(-450,25){$\boldsymbol{T=0.01}$}
\caption{Capacity of QB $K$ versus $D_z$ for the AFM case in the Heisenberg $XYZ$ model with parameters $G_z=0$, $J=1$, $\Delta=0.5$, $\gamma=0.5$, $\Omega=1$, and $\theta=\frac{\pi}{2}$. The subfigures represent different temperatures: (a) $T=0.01$, (b) $T=0.1$, and (c) $T=1$.}
\label{f10}
\end{figure*}
\begin{figure*}[!t]  
    \centering
    \includegraphics[width=0.9\textwidth]{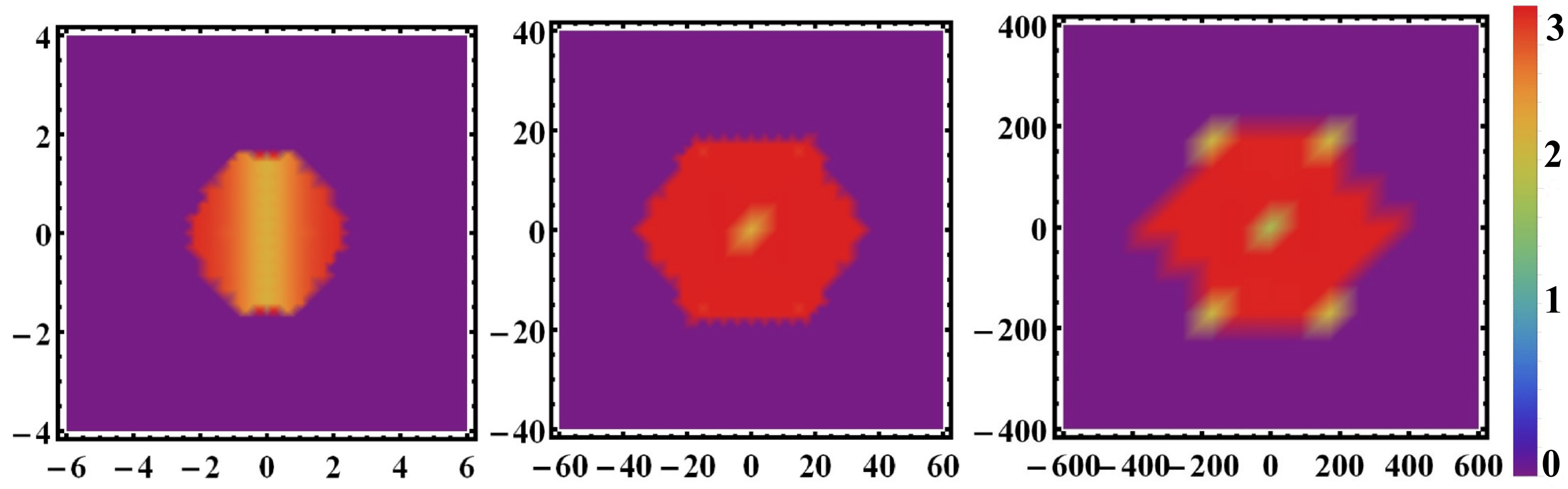}  
\put(-410,140){$\boldsymbol{(a)\, T=0.01}$}
\put(-270,141){$\boldsymbol{(b)\, T=0.1}$}
\put(-110,142){$\boldsymbol{(c)\, T=1.0}$}
\put(-384,-10){$\boldsymbol{D_z}$}
\put(-241,-10){$\boldsymbol{D_z}$}
\put(-90,-10){$\boldsymbol{D_z}$}
\put(-170,73){$\boldsymbol{G_z}$}
\put(-319,72){$\boldsymbol{G_z}$}
\put(-465,71){$\boldsymbol{G_z}$}
\put(2,60){\rotatebox{90}{$\boldsymbol{\mathcal{Q}_{\text{max}}(\hat{\varrho})$}}}
    \caption{Density plots of the maximized $l_1$-norm of quantum coherence, $\mathcal{Q}_{\text{max}} (\hat{\varrho})$, as functions of $D_z$ and $G_z$ for the AFM Heisenberg $XYZ$ model with $J = 1$, $\Delta = 0.5$, $\gamma = 0.5$, $\Omega=1$, and $\theta = \frac{\pi}{2}$. Subfigures display different temperatures: (a) $T = 0.01$, (b) $T = 0.1$, and (c) $T = 1$.}
    \label{f11}
\end{figure*}

\subsubsection{Effect of DM Coupling}
The relationship between $\xi(t)$, $\Omega t$, and $D_z$ for the AFM case with $\theta = \frac{\pi}{2}$, $G_z = 0$, and $T = 0.1$ is shown in Fig. \ref{f7}. The density plots show that variations in DM coupling do not affect the charging rate across the cases depicted in Figs. \ref{f7}(a) to \ref{f7}(d). Remarkably, increasing the DM coupling enhances ergotropy values up to a threshold near $D_z \approx 35$, consistent across all configurations from $XX$ to $XYZ$. Beyond this threshold, $\xi(t)$ suddenly drops down to zero, similar to the effect of $G_z$ on $\xi(t)$ in Fig. \ref{f6}. { This pattern, where $\xi(t)$ increases with $D_z$ up to a point and then abruptly drops to zero, persists across all AFM cases, indicating a consistent trend.

In the FM case, as shown in Fig. \ref{f8}, the variation of $\xi(t)$ with $D_z$ and $\Omega t$ reveals a similar behavior. Like the AFM scenarios where $\xi(t)$ drops suddenly after exceeding a certain threshold of $D_z$, the FM case also shows these sudden drops.}

To further extend our analysis of why this happens, we focus on the $XYZ$ case as an example of complete anisotropy. Both the FM and AFM scenarios exhibit similar behaviors in the $XYZ$ configuration, including the abrupt transition of $\xi(t)$ with changes in $D_z$. Our goal is to determine the maximum ergotropy $\xi_{max}$ achieved by the QB in the AFM scenario and investigate its variation with $D_z$ at different temperatures. We maintain the other parameters as in Figs. \ref{f7} and \ref{f8}, and perform the analysis using numerical optimization.
Various temperature values have been chosen to show the dependence of $\xi_{\text{max}}$ on $D_z$ in Figs. \ref{f9}(a)-(d).

The plots are generated under the AFM scenario with parameters $G_z=0$, $J=1$,  $\Delta=0.5$, and $\gamma=0.5$. These figures illustrate how the maximum ergotropy $\xi_{\text{max}}$ varies with $D_z$ at different temperatures.

At a low temperature such as $T=0.01$ (Fig. \ref{f9}(a)), $\xi_{\text{max}}$ initially increases with $D_z$ up to a threshold value of $D_z \approx 2.25$, then sharply drops to zero. The inset in Fig. \ref{f9}(a) shows a narrow window of oscillations before the value reaches zero, and $\xi_{\text{max}}$ does not recover for any further increase even until $D_z\rightarrow \infty$. Similar behavior is observed at higher temperatures, with $\xi_{\text{max}}$ showing larger values and a higher threshold of $D_z$ as shown in Figs. \ref{f9}(b) and \ref{f9}(c). The maximum value of $\xi_{\text{max}}$ corresponding to the threshold $D_z$ increases with temperature.

Recently, a new measure, known as the \emph{capacity} of QB is proposed \cite{yang2023battery}, defined as
\begin{equation}
K = \text{Tr}[\mathcal{\hat{H}}_{\mathcal{Q}} \hat{\varrho}_{\uparrow}] - \text{Tr}[\mathcal{\hat{H}}_{\mathcal{Q}} \hat{\varrho}_{\downarrow}],
\label{eq32}
\end{equation}
where $\hat{\varrho}_{\downarrow}$ denotes the lowest energy state of the QB, which in our case is the Gibbs thermal state defined in Eq. \eqref{eq88}, and $\hat{\varrho}_{\uparrow}$ is the highest energy state, $\hat{\varrho}_{\uparrow}=|11\rangle \langle 11|$. { It is worth noting that the ergotropy, representing the maximum extractable work from the QB, typically does not equal $K$, except in the specific cases where the temperature is zero ($T=0$) or when the vacuum state is used instead of a thermal state. The evaluation of the capacity of the QB is independent of time, relying solely on the lowest and highest energy states. This makes it applicable for evaluating both unitary and non-unitary QBs. This measure, which requires no optimization, can be linked to specific entanglement and quantum coherence measures, as demonstrated in \cite{yang2023battery}. However, its values are not directly comparable to the maximum ergotropy.}

To support the trend observed in Fig. \ref{f9} for the peak value of ergotropy versus $D_z$, we evaluate $K$ for the same parameter values used in Fig. \ref{f10}.

The closed-form expression for the capacity of QB turns out to be (with $S = \sin(\theta)+\cos(\theta)$)
\begin{equation}
K = \frac{2 \sqrt{a} f_a h+\sqrt{b} \left(d^2-1\right)+d^2 S+2 h \epsilon_a  (2 \Delta +S)+S}{d^2+2 h \epsilon_a +1}.
\end{equation}

Figure \ref{f10} shows $K$ as a function of $D_z$ at various temperatures: $T=0.01$ in Fig. \ref{f10}(a), $T=0.1$ in Fig. \ref{f10}(b), and $T=1.0$ in Fig. \ref{f10}(c). The plots are generated under the AFM scenario with fixed parameters $G_z=0$, $J=1$, $\Delta=0.5$, and $\gamma=0.5$.

These plots reveal how $K$ varies with $D_z$ at different temperatures. At $T=0.01$, $K$ initially increases with $D_z$ up to a threshold value of $D_z \approx 2.25$, and then sharply drops to zero. { The inset plot in Fig. \ref{f10}(a) highlights a narrow and sharp oscillation window before $K$ reaches zero and does not recover for any higher values of $D_z$.} Similar behavior is observed at higher temperatures, with $K$ showing larger values and a higher threshold of $D_z$ as shown in Figs. \ref{f10}(b) and \ref{f10}(c). The maximum value of $K$ corresponding to the threshold $D_z$ increases with temperature.

We observe a similarity between the capacity of the QB and the peak value of ergotropy with respect to the threshold value of $D_z$, where both measures reach their maximum and then suddenly drop to zero. Although the peak values of these quantifiers differ, they exhibit the same trend concerning the threshold value of $D_z$ at various fixed temperatures. This suggests that the QB has a finite capacity for supplying maximum work, which explains why infinite ergotropy cannot be achieved by continuously increasing $D_z$. The QB's capacity also reflects this constraint, showing that while increasing $D_z$ up to a certain threshold can enhance the QB's capacity, it cannot be extended beyond that limit.

To investigate whether the observed signature in ergotropy and QB capacity is linked to an abrupt transition in quantum coherence, { we evaluate the $l_1$-norm of coherence $\mathcal{Q(\hat{\varrho})}$ for the AFM case of $XYZ$ scenario.  Specifically, let us consider the following density matrix}
\begin{equation}\hat{\varrho}=\hat{U}_{\mathcal{C}} \hat{\varrho}_{th}\hat{U}_{\mathcal{C}}^\dagger,
\end{equation}
where $\hat{\varrho}_{th}$ and $\hat{U}_{\mathcal{C}}$ are defined in Eqs. \eqref{eq88} and \eqref{eq:unitary_operator}, respectively.

The $\mathcal{Q}(\hat{\varrho})$ is defined by \cite{baumgratz2014quantifying}
\begin{equation}
Q(\hat{\varrho}) = \sum_{i\neq j} |\langle i|\hat{\varrho}|j\rangle| = \sum_{i,j} |\hat{\varrho}_{ij}| - \sum_i |\hat{\varrho}_{ii}|,
\label{eq31}
\end{equation}
which is the sum of the absolute off-diagonal elements of a quantum state $\hat{\varrho}$ in the reference basis.

Figure \ref{f11} illustrates the variations in the maximum $l_1$-norm of quantum coherence for the AFM Heisenberg $XYZ$ model versus $D_z$ and $G_z$, using the same fixed parameter values as in Figs. \ref{f9} and \ref{f10} for comparison. Similar to the plots of $\xi_{\text{max}}$ and the QB's capacity shown in Figs. \ref{f9} and \ref{f10}, $Q_{\text{max}} (\hat{\varrho})$ displays a comparable trend. A sudden drop in $Q_{\text{max}} (\hat{\varrho})$ around the same threshold values of DM and KSEA interactions is observed, analogous to the behaviors of ergotropy and capacity. This clearly indicates that the amount of ergotropy is explicitly bounded by the capacity of the QB, and capacity is constrained by the availability of quantum coherence, implying that ergotropy, or the maximum work extraction, is ultimately (implicitly) limited by the QB's coherence.

Furthermore, an increase in temperature requires even larger values of $D_z$ to achieve the same maximum quantum coherence ($\approx 3$ as shown in Fig. \ref{f11}(a)-(c)), as also observed in the QB's capacity plot in Fig. \ref{f10}. This suggests that indefinitely extracting work from a spin-based QB by merely increasing the values of DM or KSEA interactions is not feasible. The extractable work is constrained by the QB's capacity and ultimately limited by the availability of quantum coherence.

The occurrence of unexpected oscillation followed by non-analytical behavior in $\xi_{\text{max}}$ and $K$ in Fig. \ref{f9} and Fig. \ref{f10}, especially within the insets, indicates a kind of sudden phase transition. Initially, the ergotropy $\xi(t)$ increases with $D_z$, reflecting enhanced work capacity due to quantum coherence and correlations. However, beyond a critical $D_z$, $\xi_{\text{max}}$ abruptly drops to zero, signifying a transition where the essential quantum coherence for work extraction is lost. This behavior is mirrored in the QB's capacity $K$ and the $l_1$-norm of coherence $\mathcal{Q}(\hat{\varrho})$, which also sharply decline at the same threshold. This consistent pattern suggests a fundamental reorganization of the system's quantum state, highlighting the critical role of quantum coherence in work extraction and underscoring the influence of quantum correlations on the system's energetic properties.

\section{Conclusion and outlook}\label{sec:4}
In this paper, we provided a thorough investigation of finite spins quantum batteries based on four different subcategories of Heisenberg spin chains ($XX$, $XY$, $XXZ$, $XYZ$), integrating the variable Zeeman  splitting and  DM and KSEA interactions. By exploring the maximum extractable work from diverse spin chain models in both ferromagnetic and antiferromagnetic settings, we uncovered significant insights into the mechanisms of energy storage and extraction in these quantum systems.

Our results indicate that the performance of these quantum batteries is notably affected by several factors:

\begin{itemize}
    \item \textbf{Inhomogenous Zeeman splitting and temperature}: In antiferromagnetic systems, optimal energy extraction is achieved when the external magnetic field is applied to one of the two spins. In contrast, ferromagnetic systems benefit from applying the magnetic field uniformly to both spins for certain fixed parameter values. Generally, lower temperatures enhance performance across most models. However, the $XY$ model in antiferromagnetic configurations exhibits considerably improved robustness at elevated temperatures. Conversely, in ferromagnetic systems, energy extraction diminishes with increasing temperatures due to heightened thermal fluctuations.

    \item \textbf{Symmetric-antisymmetric exchange interactions}: The incorporation of specific spin interactions, such as Dzyaloshinsky--Moriya and Kaplan--Shekhtman--Entin-Wohlman--Aharony couplings, can substantially boost the energy storage and extraction capabilities of these quantum batteries. Nevertheless, there exists a threshold beyond which further increases in interaction strength result in a sharp decline in energy storage capacity.

    \item \textbf{Model-specific behavior}: Different Heisenberg spin chain models ($XX$, $XY$, $XXZ$, $XYZ$) exhibit distinct behaviors under varying conditions, highlighting the importance of model selection for optimizing battery performance in specific applications.

    \item \textbf{Quantum phenomena}: The interplay between various quantum phenomena, including coherence and entanglement, significantly influences the performance of the battery. The maximum amount of extractable energy is limited by both the battery's inherent capacity and the level of quantum coherence present within the system.
\end{itemize}

These findings highlight the interplay between energy extraction, quantum battery capacity, and quantum coherence under varying conditions. Our findings reinforce the theoretical framework for finite spin chain-based quantum batteries and provide practical insights into material selection and experimental setups.
By recognizing the limitations of quantum resources, this study underscores the need for strategies to optimize their use in advanced quantum systems. Moreover, one can improve the efficiency of energy storage and extraction in quantum batteries by controlling magnetic fields, symmetric and antisymmetric exchange interactions, temperature, and quantum effects.

In summary, this research offers valuable insights into optimizing quantum batteries derived from spin systems, contributing to the development of efficient energy storage devices in quantum technologies.

\section*{Acknowledgements}
M.G. and S.H. were supported by Semnan University under Contract No. 21270.

\section*{Disclosures}
The authors declare that they have no known competing financial interests.

\bibliography{bibliography}

\end{document}